\documentclass[a4paper,fleqn]{cas-sc}

\usepackage[authoryear]{natbib}
\usepackage{subfigure}
\usepackage{threeparttable}
\usepackage{float}
\usepackage{multirow}
\usepackage{booktabs}
\usepackage{setspace}
{}
{}
{}
\usepackage{graphicx}

\begin{document}
\let\WriteBookmarks\relax
\def\floatpagepagefraction{1}
\def\floatpagepagefraction{2}
\shorttitle{METcross: A framework for short-term forecasting of cross-city metro passenger flow}
\shortauthors{}

\title [mode = title]{METcross: A framework for short-term forecasting of cross-city metro passenger flow}                      

\address[1]{School of Transportation, Southeast University, Nanjing 211189, China}
\address[2]{College of Transportation Engineering, Chang’an University, Xi’an 710064, China}
\address[3]{School of Traffic and Transportation, Beijing Jiaotong University, Beijing 100091, China}
\author[1]{Wenbo Lu}[style=chinese]
\author[2]{Jinhua Xu}[style=chinese]
\author[3]{Peikun Li}[style=chinese]
\author[1]{Ting Wang}[style=chinese]
\author[1]{Yong Zhang}[style=chinese]
\cormark[1]
\cortext[cor1]{School of Transportation, Southeast University, Nanjing 211189, People's Republic of China}
\cortext[cor1]{E-mail: zhangyong@seu.edu.cn}

\begin{abstract}
Metro operation management relies on accurate predictions of passenger flow in the future. This study begins by integrating cross-city (including source and target city) knowledge and developing a short-term passenger flow prediction framework (METcross) for the metro. Firstly, we propose a basic framework for modeling cross-city metro passenger flow prediction from the perspectives of data fusion and transfer learning. Secondly, METcross framework is designed to use both static and dynamic covariates as inputs, including economy and weather, that help characterize station passenger flow features. This framework consists of two steps: pre-training on the source city and fine-tuning on the target city. During pre-training, data from the source city trains the feature extraction and passenger flow prediction models. Fine-tuning on the target city involves using the source city's trained model as the initial parameter and fusing the feature embeddings of both cities to obtain the passenger flow prediction results. Finally, we tested the basic prediction framework and METcross framework on the metro networks of Wuxi and Chongqing to experimentally analyze their efficacy. Results indicate that the METcross framework performs better than the basic framework and can reduce the Mean Absolute Error and Root Mean Squared Error by 22.35\% and 26.18\%, respectively, compared to single-city prediction models.
\end{abstract}
\maketitle

\section{Introduction}
The development of metro as a solution to alleviate urban road congestion has been rapid. Due to the low carbon emission and punctuality, metro is increasingly popular among urban travelers. Accurate and real-time short-term passenger flow (STPF) prediction is essential to daily metro operations and offer a useful reference for management departments. However, the complexity of urban travel environments makes STPF tasks challenging.

Researchers have developed various prediction models from multiple perspectives, ranging from feedforward neural network models such as MLP (Multilayer Perceptron) to recurrent neural network models like LSTM (Long Short-Term Memory), that often model individual stations \citep{xia2013passenger};\citep{yang2021novel}; \citep{zhang2023comparative}. In recent years, GNN (Graph Neural Network) models, which can consider the correlation between multiple stations, have been rapidly developed and widely used \citep{han2019predicting}; \citep{zhang2020multi}; \citep{xu2023multi}. Furthermore, GNN models based on multi-view modeling have been a hotspot of research \citep{jin2021hetgat}; \citep{lu2021dual}; \citep{wu2023learning}. Deep learning-based models have been shown to have the smallest errors in STPF prediction \citep{liu2019deeppf}. 

While significant progress has been made in model construction in previous studies, relatively few have adopted a cross-city knowledge integration approach, which has the potential to limit the development of prediction models. In reality, modeling cross-city STPF prediction can utilize passenger flow data from different cities, and using passenger flow data from the source city can enhance capturing the trend of passenger flow variations. For instance, two stations in different cities may share similar passenger flow characteristics where using the passenger flow data from the source city can enhance modeling. This can be demonstrated through the prediction of traffic flow in urban areas \citep{wang2018cross}.

We aim to use metro passenger flow data from the source city to assist in predicting target city passenger flow. However, cross-city passenger flow prediction encounters challenges such as: 1.How to utilize passenger flow data from the source city to improve prediction modeling for the target city? 2.How best to represent the passenger flow characteristics of different metro stations in different cities and determine the input data elements?
3.How to standardize the dimensions of input data due to the urban metro network's scale differences between the source and target cities? 4.How to prevent the use of harmful knowledge from the source city that could negatively impact cross-city passenger flow prediction?

Based on the issues mentioned, our considerations are as follows:
\begin{itemize}
\item Firstly, we constructed a basic framework for cross-city metro STPF prediction, incorporating data fusion and transfer learning perspectives.
\item Next, we comprehensively constructed static and dynamic covariates from the perspectives of economy, transportation, geographical location, and weather, to represent the passenger flow characteristics of stations.
\item Furthermore, we utilized the correlation between the passenger flow sequences of source city and target city metro stations to develop adjacency matrices, weight matrices, and similarity matrices as methods for dimension transformation.

\item Finally, we utilized the static and dynamic covariates to build the cross-city metro STPF prediction framework METcross which includes residual connections to guarantee that the prediction error is at least no larger than that of single-city prediction modeling.
\end{itemize}

In this study, we calculated the Pearson coefficient to measure the passenger flow sequences correlation between metro stations in the source and target cities. Subsequently, we construct adjacency matrices, weight matrices, and similarity matrices between the two cities' metro stations. Using these matrices as methods for data dimension transformation, we establish the basic framework for cross-city metro STPF prediction from the perspectives of data fusion and transfer learning.

Furthermore, we construct static and dynamic covariates considering network-level factors like weather and economy, and station-level factors like functional mixture and geographical location. We utilize these covariates as inputs to construct the METcross prediction framework, which includes the pre-training and fine-tuning procedures. We use an encoder-decoder network in the pre-training process to encode and decode the input data from the source city to obtain feature embeddings. In the next step, a prediction network generates predictions for the source city and is trained. In the fine-tuning process, we transfer the pre-trained encoder-decoder network to the target city. Feature embeddings from the source and target cities are fused to attain initial predictions. Additionally, we model the passenger flow data from the target city to acquire baseline predictions. Finally, we use residual connections to merge the initial and baseline predictions to produce the final  prediction results.

To summarize, this study offers the following contributions:
\begin{itemize}
\item Firstly, we provide a basic framework for cross-city metro STPF prediction modeling, using input data, feature, and prediction result fusion, along with transfer learning.

\item Secondly, we analyze the prediction errors resulting from different fusion methods and transformation techniques, offering a theoretical basis for joint modeling of cross-city metro passenger flow prediction.

\item Thirdly, we construct the METcross framework for cross-city metro STPF prediction. It's worth noting that our study represents the first attempt to use source city metro passenger flow data to aid STPF prediction tasks for the target city.

\item  Lastly, we conduct an exhaustive experiment on real metro datasets from two cities to validate the effectiveness of our proposed framework.
\end{itemize}

This paper is structured as follows: Section 2 elaborates on the STPF prediction research. Section 3 introduces the relevant fundamental concepts, modeling objectives, and notation definitions for the study. Section 4 depicts the basic and METcross framework, including their construction. In section 5, the application results of both the basic and METcross prediction frameworks are presented with a focus on two cities. Lastly, section 6 concludes the study.
\section{Literature Review}
STPF in metro is a common time series prediction task that has been explored extensively by researchers. Numerous prediction models have been developed from different perspectives, such as data preprocessing, multi-source data fusion, model construction, and modeling scenarios.
\subsection{Decomposition and aggregation modeling}
From the data preprocessing standpoint, the decomposition and aggregation of passenger flow data constitutes a widely adopted technique \citep{li2023traffic}; \citep{tang2022seasonal}. Data decomposition yields multiple sub-sequences that can be modeled independently, thereby facilitating the analysis of passenger flow fluctuations with distinctive characteristics. Decomposition methods such as Empirical Mode Decomposition and Variational Mode Decomposition decompose passenger flow time series into components that represent features at various time scales or frequency ranges \citep{huang2023deaseq2seq}; \citep{liu2020short}; \citep{wei2012forecasting}; \citep{zhang2020lightgbm}. Another popular method is Seasonal and Trend decomposition using Loess, which dissects the passenger flow sequence into trend, seasonality, and residuals. Decomposition outcomes are more lucid and easier to comprehend. However, it may struggle to handle nonlinearity, non-stationarity, or complex data structures \citep{chen2020forecasting}; \citep{qin2019effective}.

Alternatively, the aggregation method involves combining multiple original passenger flow data, thereby facilitating information sharing among similar passenger flow sequences \citep{lu2023mohp}; \citep{sajanraj2021passenger}. To group stations and learn feature embeddings for similar station, researchers make use of clustering algorithms. For instance, \citep{wei2022cluster} employed clustering algorithms to partition all metro stations into several groups so that members of each group could learn shared embeddings. Modeling the correlation among functional areas of each station using an adjacency matrix provides more accurate spatial information for prediction. Another study \citep{tu2022forecasting} suggests that the functional type of metro stations can be determined by their surrounding Points of Interest (POI). The researchers employed POI to calculate the functional types of different metro stations and developed a deep learning architecture called DeepSPF to forecast metro flow at stations with varying functional types.
\subsection{Multiple-source data fusion}
Integrating various external factors that impact metro passenger flow can significantly enhance the model's capability from the perspective of fusing multiple source data. Weather, for instance, is a significant determinant of travel behavior, and is widely used as an additional input. \citep{zhang2020deep} took into account weather conditions and air quality and fused them with residual networks (ResNet), graph convolutional networks (GCN), and LSTM to predict metro passenger flow. Furthermore, the land-use characteristics around metro stations play a crucial role in determining the flow patterns. \citep{zeng2023combining} proposed a metro knowledge graph construction method that incorporated land-use features to adapt to the metro system's prior understanding. Additionally, mobile signaling data, which reflect traveler activities, are frequently employed in metro STPF prediction modeling. To achieve online estimation of passenger flow, \citep{tao2021online} utilized automatic differentiation to fuse multiple data sources, such as AFC data, mobile signaling data, and historical passenger flow data.

Another useful data is search engine data, which can illustrate current hot events and their impact on travel behavior. \citep{jin2022novel} collected data from Baidu's information-rich search index, reduced its dimensionality, and selected potent predictors through statistical analysis. They put forth a novel hybrid model that incorporates the Baidu search index for multi-step ahead prediction of metro passenger flow. Furthermore, \citep{li2022metro} fused related indexes from multiple sources, obtained from three major search engines (Baidu, Sogou, 360), as additional inputs, and devised a new method to combine multiple-source time series for STPF prediction.
\subsection{Methods based on deep learning}
In the realm of metro STPF prediction, deep learning models are particularly powerful due to their capabilities for nonlinear modeling. LSTM models, in particular, have gained popularity for predicting single-station passenger flows, as they can effectively capture both long-term and short-term temporal features present in passenger flow sequences \citep{hao2019sequence}. Moreover, introducing the correlation among stations has significantly improved prediction performance. To this end, CNNs (Convolutional Neural Networks) are often used because they can capture spatial correlations among stations \citep{ma2018parallel}. Although CNNs have an inherent ability to capture local perceptions, they frequently face challenges while detecting global information. Furthermore, metro networks ought to be treated as graph-structured data, necessitating the presence of global topology and relationships to improve the task's accuracy. GNNs are particularly useful for this purpose, as they can establish connections among nodes and effectively propagate information globally. The graph convolutional layer in GNN models the relationships between every pair of stations, and subsequently shares the information with the neighboring stations to enhance their passenger flow representations. Therefore, for every station, GNN allows it to enrich its passenger flow representation by leveraging information obtained from neighboring stations.

Constructing graphs and developing relationships between stations is a critical step for utilizing GNN in STPF prediction. Typically, the physical metro network serves as the basis for constructing a typical graph structure. The relationships between stations are established based on their physical topological connections. However, this approach may not adequately reflect the actual spatial dependencies between stations. Recent studies have proposed alternative methods for improving the graph structure in metro STPF prediction. For instance, \citep{zhao2023adaptive} utilized a trainable adaptive adjacency matrix in constructing an adaptive graph convolutional network model, which showed superior performance over models that used fixed adjacency matrices. Additionally, \citep{wang2021metro} introduced hypergraphs, which enable the extraction of high-order relationships between stations and passengers' travel patterns, leading to the development of a dynamic spatiotemporal hypergraph neural network that can predict metro passenger flow.

GNN prediction models based on multi-view fusion are widely used to integrate relationships between stations from various perspectives. To capture the underlying relationships between stations, \citep{li2023ig} developed three types of inter-station interaction graphs, including connectivity, similarity, and temporal correlation graphs, to simulate interactions between stations from different points. \citep{bao2022forecasting} introduced spatial heterogeneity correlations between stations by examining geographical distance view, functional similarity view, and demand pattern view, respectively. \citep{liu2020physical} integrated several graph structures, including the physical metro network, passenger flow similarity graph, and correlation graph, into a graph convolutional gated recurrent unit for effective spatiotemporal representation learning. \citep{wang2023network} explored three similarity views (i.e., adjacency similarity, geographical location similarity, and trend similarity) for developing a more accurate STPF prediction model. Moreover, as the relationships between metro stations may change dynamically in practice, there is a need to learn dynamic spatial relationships between metro stations for accurate prediction \citep{xie2023spatio}.
\subsection{Multi-scenario modeling}
From the perspective of modeling scenarios, prediction models constructed for different scenarios such as holidays and special events can achieve better results. Holidays, due to their suddenness and irregularity, pose more challenges for STPF prediction tasks. \citep{zhang2023spatiotemporal} combined GNN and attention mechanism to predict metro passenger flow during the New Year holiday. By integrating social media data, they comprehensively captured the influence of social media on holiday passenger flow and enhanced understanding of the evolving trends of holiday passenger flow. \citep{wen2022decomposition} decomposed the time series into linear and nonlinear components to identify holiday factors. In addition, transfer learning was used to predict holiday passenger flow, while the SARIMA(Seasonal Autoregressive Integrated Moving Average) model was used to predict regular passenger flow.

While predicting passenger flow on weekdays and holidays is important for normal metro operations, special events such as concerts and football matches that result in significant passenger flow require much attention from managers. \citep{guo2019short} developed a fusion model for detecting and predicting anomalous passenger flow using support vector machine and LSTM. Similarly, \citep{ni2016forecasting} explored the relationship between social media data volume and passenger flow during large events. Twitter user data was leveraged to predict the total passenger flow for the next four hours. In addition, \citep{chen2019subway} proposed a generalized autoregressive conditional heteroskedasticity model to predict the mean and volatility of passenger flow during events and forecast passenger flow for the next ten minutes.

Merely predicting passenger flow during special events does not provide sufficient time for management departments to take measures. To address this issue, \citep{wang2019early} proposed a model that simulates the evacuation process of additional outbound passenger flow after an event using Newton's cooling law, which can predict the occurrence of large passenger flows caused by special events up to two hours in advance. Additionally, \citep{lu2023estimating} integrated various data sources, including passenger flow information during special events, station information, and event information, to construct a model for estimating large passenger flows of special events.

In conclusion, although scholars have developed STPF prediction models from multiple perspectives, the majority of these models have focused on modeling passenger flow at individual metro stations within a single city. Consequently, there has been limited research on cross-city passenger flow prediction tasks. However, utilizing abundant passenger flow data from multiple cities could make it possible for breaking the bottleneck of prediction models.
\section{Preliminaries}
\subsection{Notations}
\noindent \textbf{Definition 1 (Metro Network)}: The sets of stations in the source and target city are denoted as $\boldsymbol{S}=\left\{ r^1,r^2,...,r^s,...,r^S \right\}$ and $
\boldsymbol{G}=\left\{ r^1,r^2,...,r^g,...,r^G \right\}$, respectively, with $S$ and $G$ representing the number of stations.

\noindent \textbf{Definition 2 (Metro Passenger Flow)}: Metro passenger flow exhibits both temporal and spatial characteristics. The set of time periods is defined as $
T=\left\{ 1,2,...,t,...,T \right\}$. The passenger flow of source city station $r^s$ and target city station $r^g$ in time period $t$ is denoted as $
x_{t}^{s}$ and $x_{t}^{g}$; The length $T$ passenger flow sequences are represented as $
\boldsymbol{x}_{T}^{s}=\left[ x_{1}^{s},x_{2}^{s},...,x_{T}^{s} \right] 
$ and $\boldsymbol{x}_{T}^{g}=\left[ x_{1}^{g},x_{2}^{g},...,x_{T}^{g} \right] $.

\noindent \textbf{Definition 3 (Input Features)}: According to the literature \citep{zhang2019multistep}, the input features consist of historical passenger flows for the previous $h$ time periods and their averages. Thus, the passenger flow input matrix for the source city in time period $t$ is denoted as $
\mathbb{L}_{t}^{\boldsymbol{S}}=\left[ \boldsymbol{L}_{t}^{1};\boldsymbol{L}_{t}^{2};...;\boldsymbol{L}_{t}^{s};...;\boldsymbol{L}_{t}^{S} \right] \in \mathbb{R}^{S\times \left( h+1 \right)}
$, $\boldsymbol{L}_{t}^{s}=\left[ x_{t-h}^{s},x_{t-h+1}^{s},...,x_{t}^{s},\bar{x}_{t}^{s} \right] $. In addition, the matrix of static covariate features is denoted as $\mathcal{A}_{t}^{\boldsymbol{S}}=\left[ \boldsymbol{A}_{t}^{1};\boldsymbol{A}_{t}^{2};...;\boldsymbol{A}_{t}^{s};...;\boldsymbol{A}_{t}^{S} \right] \in \mathbb{R}^{S\times A}$, where $\boldsymbol{A}_{t}^{s}=\left[ p_{t}^{1},p_{t}^{1},...,p_{t}^{a},...,p_{t}^{A} \right]$; $p_{t}^{a}$ represents the $a-th$ static covariate, such as metro station function information, etc. The dynamic covariate features is represented by $\mathcal{D}_{t}^{\boldsymbol{S}}=\left[ \boldsymbol{D}_{t}^{1,\boldsymbol{S}};\boldsymbol{D}_{t}^{2,\boldsymbol{S}};...;\boldsymbol{D}_{t}^{d,\boldsymbol{S}};...;\boldsymbol{D}_{t}^{D,\boldsymbol{S}} \right] \in \mathbb{R}^{D\times \left( h+1 \right)}$, where $
\boldsymbol{D}_{t}^{d,\boldsymbol{S}}=\left[ q_{t-h}^{d,\boldsymbol{S}},q_{t-h+1}^{d,\boldsymbol{S}},...,q_{t}^{d,\boldsymbol{S}},...,\bar{q}_{t}^{d,\boldsymbol{S}} \right]$; $q_{t}^{d,\boldsymbol{S}}$ denotes the $d-th$ dynamic covariate in time period $t$, such as temperature, air quality, etc. The set of input feature matrices for the source city is denoted as $\mathbb{H}_{t}^{\boldsymbol{S}}$. Similarly, the passenger flow input for the target city is denoted as $\mathbb{L}_{t}^{\boldsymbol{G}}\in \mathbb{R}^{G\times \left( h+1 \right)}$, the static covariate is $\mathcal{A}_{t}^{\boldsymbol{G}}\in \mathbb{R}^{G\times A}$, the dynamic covariate is $
\mathcal{D}_{t}^{\boldsymbol{G}}\in \mathbb{R}^{G\times \left( h+1 \right)}$, and the set of input feature matrices is $\mathbb{H}_{t}^{\boldsymbol{G}}$.

\noindent \textbf{Definition 4 (Station Similarity Measurement)}: The similarity matrix between two stations from the source and target city is denoted as $Si^{\boldsymbol{G}\rightarrow \boldsymbol{S}}\in \mathbb{R}^{G\times S}$, where $
Si^{g,s}=P\left( \boldsymbol{x}_{T}^{g},\boldsymbol{x}_{T}^{s} \right)$; $P\left( \cdot \right)$ represents the Pearson coefficient between the two passenger flow sequences \citep{cohen2009}. The adjacency matrix $AJ^{\boldsymbol{G}\rightarrow \boldsymbol{S}}\in \mathbb{R}^{G\times S}$ represents the relationship between stations in the target city and the source city, satisfying the following condition:
\begin{align}
AJ^{g,s}=\left\{ \begin{array}{l}
	1,\ Si^{g,s} \ge Si^{g,s*},\ \forall s^*\in \boldsymbol{S}\\
	0,\ else\\
\end{array} \right. 
\label{eq1}
\end{align}

Furthermore, the weight matrix is defined as the element-wise product of the similarity matrix and adjacency matrix, denoted as $We^{\boldsymbol{G}\rightarrow \boldsymbol{S}}=Si \circ AJ$.

\noindent \textbf{Definition 5 (Paired Stations)}: The most similar stations between the source city and target city are defined as paired stations. It should be noted that the paired stations do not necessarily form one-to-one relationships.

\noindent \textbf{Definition 6 (Model Representation)}: This study models the prediction of all metro station passenger flow. The STPF prediction model for the source city is denoted as $
F_{\theta ^{\boldsymbol{S}}}$, where $\theta ^{\boldsymbol{S}}$ represents the parameters of the model. Similarly, the STPF prediction model for the target city is denoted as $
F_{\theta ^{\boldsymbol{G}}}$.
\subsection{Modeling Motivation and Objective}
\begin{figure}[ht]
	\centering
	\includegraphics[width=10cm]{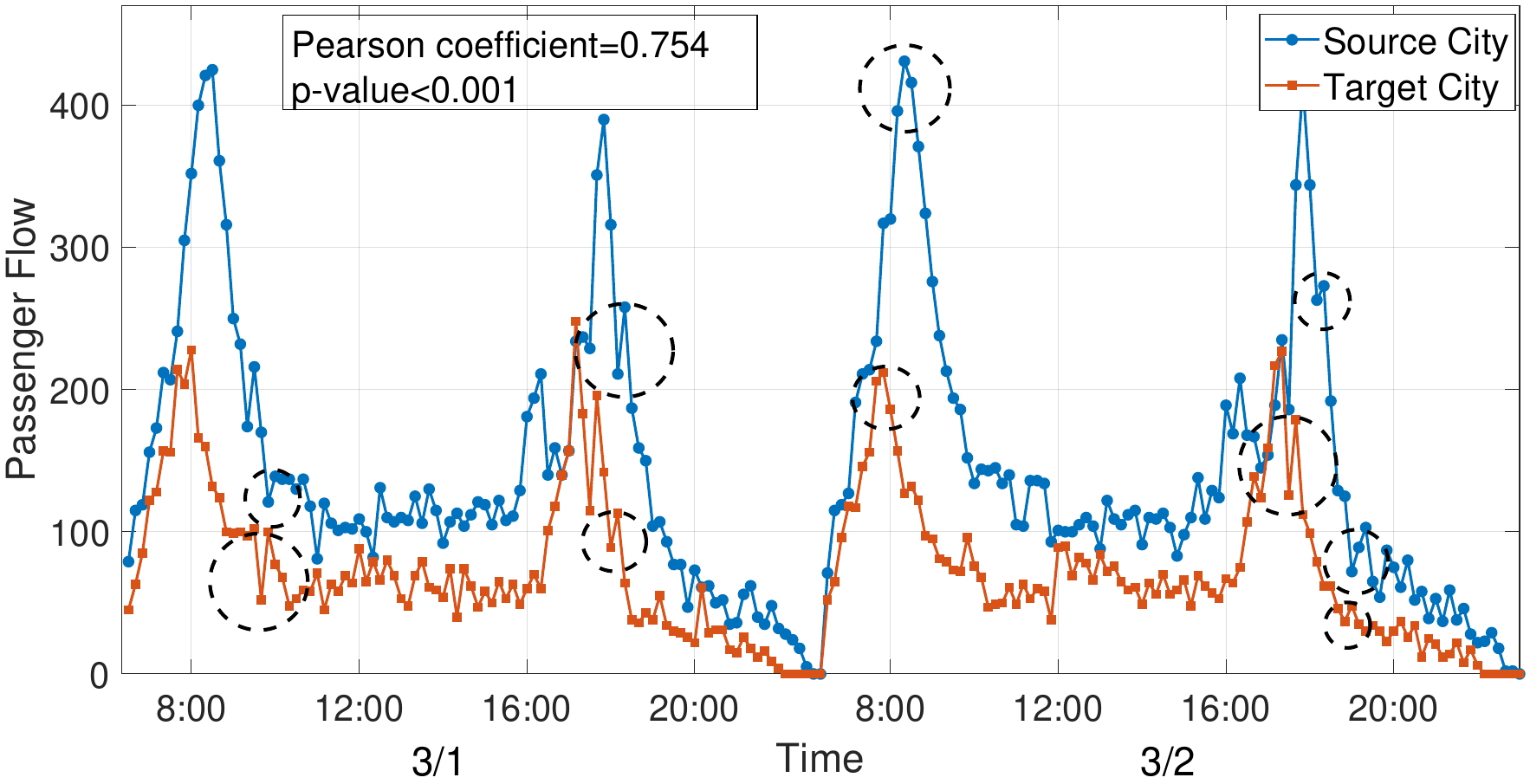}
	\caption{Pairing of passenger flow between source city and target city.}
	\label{lu1}
\end{figure}

Various factors, such as weather, economic activities, and transportation connectivity, influence the magnitude and characteristics of passenger flow at metro stations in a city. When factors like the surrounding economy and accessibility of metro stations in two cities are similar, similarities in flow patterns can be expected. Figure~\ref{lu1} displays the passenger flow time series of stations from both cities, with black circles indicating the flow pattern similarities. As a result, we utilize data from the source city to aid in predicting passenger flow in the target city.

Using input feature matrix sets for the source and target city, the objective is to use the prediction model from the source city to aid in predicting passenger flow for the target city. The goal is to reduce prediction error for the target city:
\begin{subequations}
\begin{align}
&\underset{\theta ^{\boldsymbol{G}}}{\min}\,\,\mathbb{L}\left( \mathbb{H}_{T}^{\boldsymbol{S}},\mathbb{H}_{T}^{\boldsymbol{G}},\theta ^{\boldsymbol{S}},\theta ^{\boldsymbol{G}} \right) =\sum_{t=1}^T{\sum_G{L\left( x_{t}^{g},\hat{x}_{t}^{g} \right)}}, \\
&\hat{x}_{t}^{g}=F_{\theta ^{\boldsymbol{S}},\theta ^{\boldsymbol{G}}}\left( \mathbb{H}_{T}^{\boldsymbol{S}},\mathbb{H}_{T}^{\boldsymbol{G}} \right) 
\end{align}
\end{subequations}

where, $\hat{x}_{t}^{g}$ is the predicted passenger flow at the target city station $r^g$, and $L\left( \cdot \right) $ is the error function.

\subsection{Fine-tuning for passenger flow prediction task}
Existing literature has tackled the issue of transfer learning in traffic prediction through fine-tuning, which consists of two steps. The first step entails training a model on source dataset, dubbed the pre-training process. Subsequently, the model's pre-trained parameters are employed as initial values for training on target dataset. In the particular problem under investigation in this study, the ensuing steps can be defined as follows:
\begin{subequations}
\begin{align}
&\theta ^{\boldsymbol{S}}=\min\text{\,\,}\mathbb{L}\left( \mathbb{H}_{T}^{\boldsymbol{S}},\theta \right) =\sum_{t=1}^T{\sum_{\boldsymbol{S}}{L\left( x_{t}^{s},\hat{x}_{t}^{s} \right)}}\label{3a}\\
&\theta ^{\boldsymbol{G}}=\min\text{\,\,}\mathbb{L}\left( \mathbb{H}_{T}^{\boldsymbol{G}},\theta ^{\boldsymbol{S}},\theta \right) =\sum_{t=1}^T{\sum_{\boldsymbol{G}}{L\left( x_{t}^{g},\hat{x}_{t}^{g} \right)}}\label{3b}
\end{align}
\end{subequations}

where $\theta $ represents the random initial parameters of the network, and $\hat{x}_{t}^{s}$ is the predicted passenger flow at the source city station $r^s$. Equation~(\ref{3a}) indicates the pre-training stage, while in Equation~(\ref{3b}), the parameters corresponding to $
\theta ^{\boldsymbol{S}}$ are kept unchanged during the training process.

\section{Methodology}
\subsection{Basic Framework}
We construct basic framework for both single-city and cross-city STPF prediction from five perspectives: No Fusion (NF), Input data Fusion (DF), Feature Fusion (FF), Prediction Fusion (PF), and Transfer Learning (FT).

(1) NF: Utilizes only the passenger flow data of the target city for both training and testing tasks.
(2) DF: Integrates input data from paired stations between the source and target city.
(3) FF: Combines the embedded features of paired stations between the source and target city.
(4) PF: Merges prediction results from paired stations between the source and target city.
(5) FT: Uses the passenger flow data from the source city to create a prediction network (FT-P) and feature network (FT-F) for pre-training. The parameters of the pre-trained model are used as initial values for training in the target city.

\begin{figure}[ht]
	\centering
	\includegraphics[width=13cm]{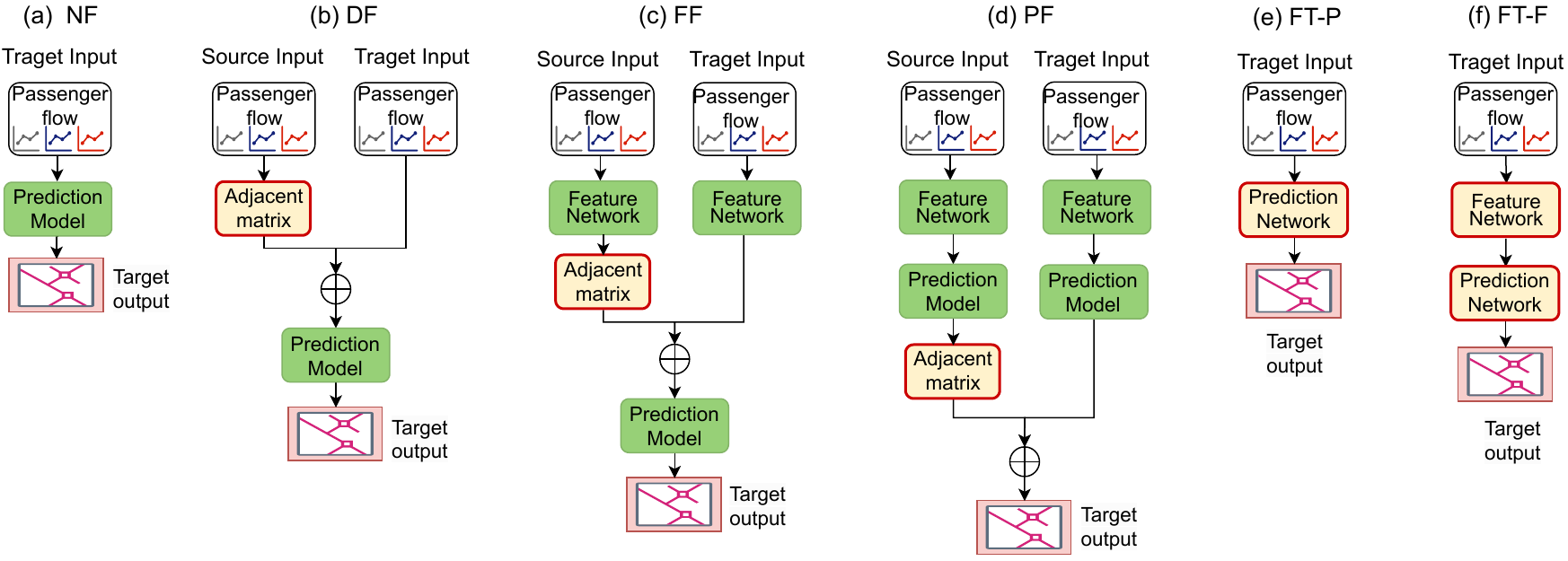}
	\caption{Basic framework for predicting passenger flow in the target city.}
	\label{lu2}
\end{figure}

The scale of metro networks in different cities causes dimensional differences that require transformation methods for alignment. Figure~\ref{lu2} illustrates the process of constructing different fusion models using the adjacency matrix as an example of the transformation method. The yellow modules represent the pre-trained or known components.
Taking DF as an example, the mathematical process is as follows:
\begin{align}
\boldsymbol{\hat{x}}_{t}^{\boldsymbol{G}}=Pre\left( AJ^{\boldsymbol{S}\rightarrow \boldsymbol{G}}\cdot \mathbb{L}_{t}^{\boldsymbol{S}}+\mathbb{L}_{t}^{\boldsymbol{G}} \right)
\end{align}

Where $\boldsymbol{\hat{x}}_{t}^{\boldsymbol{G}}\in \mathbb{R}^{G\times 1}$ is the predicted passenger flow of the target city, and $Pre\left( \cdot \right)$ is the prediction model with an input dimension of $h+1$ and an output dimension of 1.

\begin{figure}[ht]
	\centering
	\includegraphics[width=12cm]{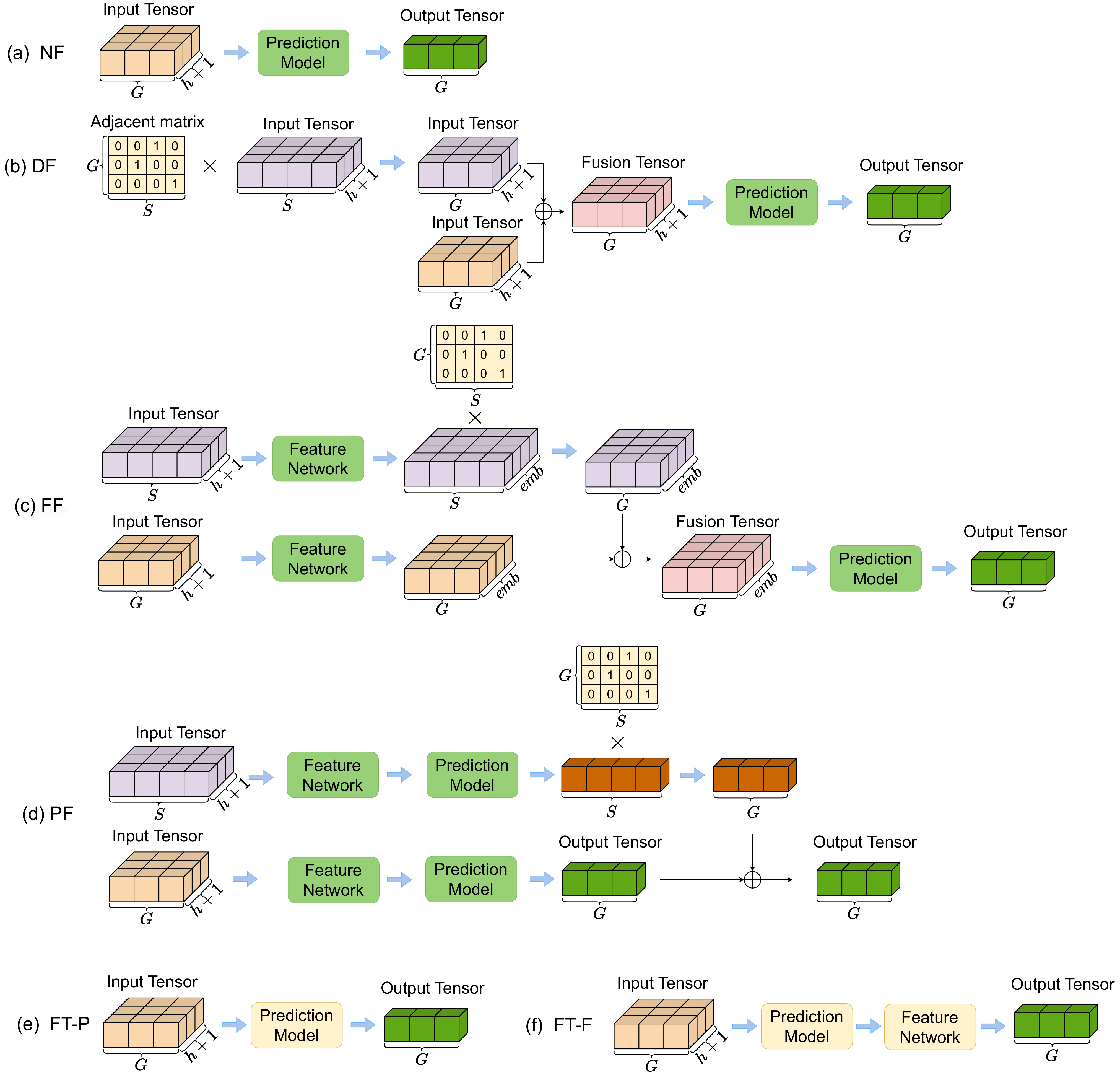}
	\caption{Process of different fusion methods.}
	\label{lu3}
\end{figure}

Furthermore, we describe the fusion process of the basic framework from the perspective of data tensors in Figure~\ref{lu3}. Due to the difference in the numbers of metro stations between the two cities, data transformation is necessary during the fusion process. We use the adjacency matrix to change the dimensions of the source city's input data, features, and prediction results to align with the target city. Additionally, weight matrix and similarity matrix are proposed as alternative transformation methods.
\subsection{METcross framework}
\begin{figure}[ht]
	\centering
	\includegraphics[width=11cm]{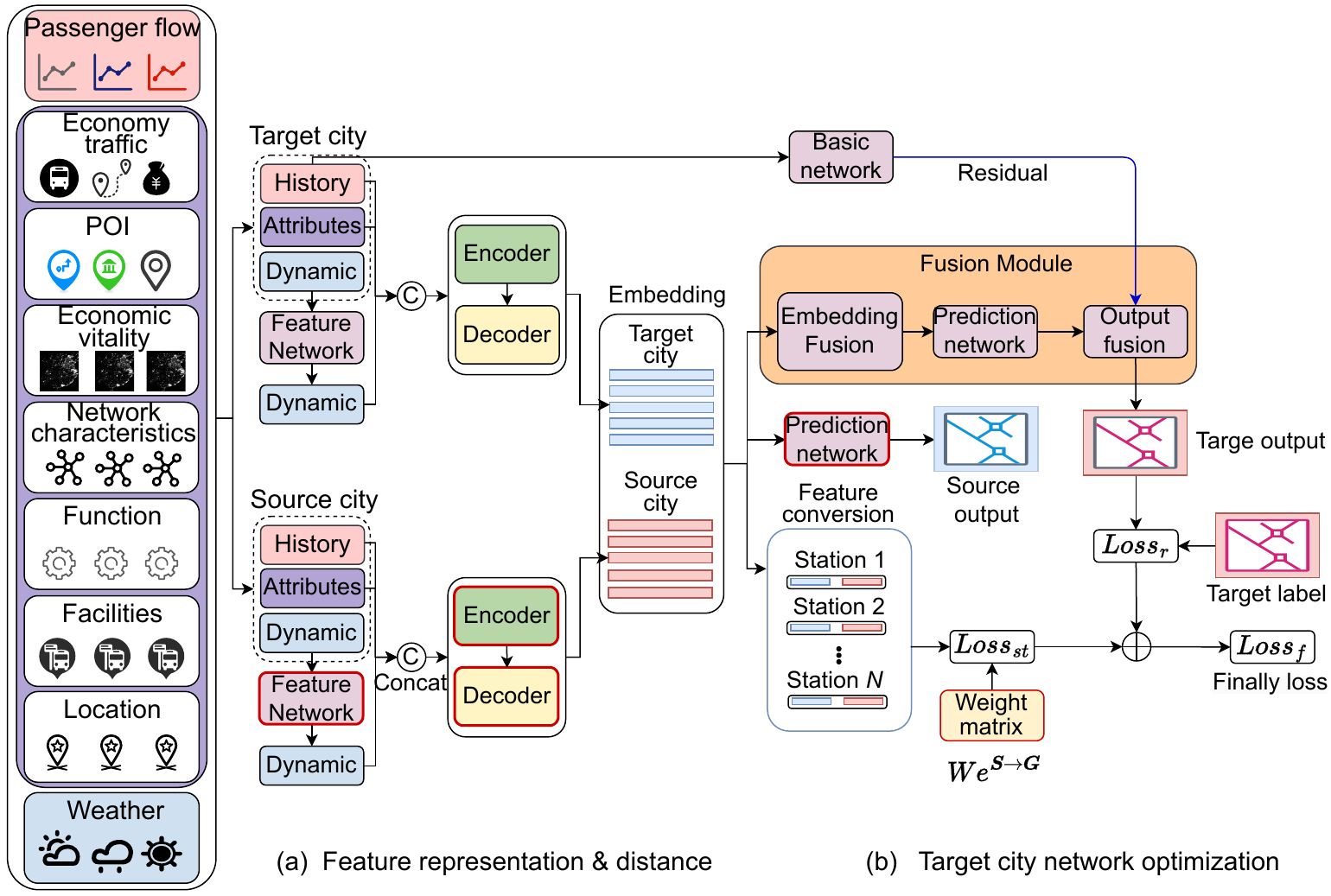}
	\caption{Overview of METcross(The red box represents the pre-training model).}
	\label{lu4}
\end{figure}

In contrast to the basic framework, the METcross framework considers both static and dynamic covariates simultaneously. We use the Encoder-Decoder structure to encode and decode features. In addition, the similarity of passenger flow feature embeddings between paired station is considered to improve model generalization. Figure~\ref{lu4} shows that the METcross framework comprises two stages and two components. During the pre-training stage, an Encoder-Decoder module is utilized to extract feature embeddings from the source city. By using Encoder-Decoder, we can convert input data and remove redundant information while retaining the main information. At the same time, the Decoder can use the features extracted by the encoder to restore the input data, thereby generating high-quality output. Furthermore, the feature embeddings are input to the prediction network to obtain predicted passenger flow and generate error signals. The parameter of each module are trained through back propagation by using the error signals.

During the fine-tuning stage, the pre-trained Encoder-Decoder is used to obtain feature embeddings for the target city. In addition, the feature embeddings of passenger flow obtained through the feature network at paired stations in the source city and target city are similar. That is, the distance between the passenger flow feature embeddings of paired stations in the source city and target city should be minimized to improve generalization.

Furthermore, the feature embeddings of both the source and target city are used in the prediction process. Firstly, the historical passenger flow of the target city is utilized to derive the baseline predicted passenger flow. Then, in the fusion module, the source city's feature embedding is merged with those of the target city to produce the initial predicted passenger flow via the prediction network. The final predicted passenger flow is arrived at by enacting a residual connection amid the initial predicted passenger flow of the source city and the baseline predicted passenger flow. The residual connection ensures that the baseline prediction value for the target city remains intact. Lastly, we integrates the prediction loss and feature loss for the sake of back propagation and updates.

The METcross framework entails two fundamental components - feature embedding distance and target city prediction error. In light of this, the framework's loss function merges the feature embedding loss with the prediction error loss, and weights them accordingly. Back propagation is then performed on the loss, and this serves to facilitate the learning of the feature extraction module, transfor network, and prediction network parameters. The framework is essentially guided by three core operations - feature representation, feature embedding distance measurement, and target city network optimization.

\subsubsection{Representation of Spatio-temporal Features}
The initial step is the extraction of feature embeddings from both the source and target city. Prior studies reveal that the factors affecting metro passenger flow can be broadly partitioned into two levels: network and station level \citep{lu2022measuring}. The network-level factors involved in this realm are weather, economy, and transportation. The station-level factors incorporate the blend of local functions, economic, reachability, station functionality, and geographic location. These factors can be further classified into static and dynamic covariates as displayed in Table~\ref{tab1}.
\begin{table}[ht]
  \scriptsize
   \setlength{\tabcolsep}{0.5mm}
   \centering
   \caption{Explanation of input features}
    \begin{tabular}{cccc}
    \toprule
    Category & Name & Variable & Explanation \\
    \midrule
    \multirow{9}[12]{*}{\makecell[c]{Static\\Covariates}} & Economy & \makecell[c]{Population, Per Capita GDP,\\ Population Density} & \makecell[c]{The more developed the urban economy, \\the greater the metro passenger flow}. \\
   \cmidrule{2-4}
   & \multirow{2}[4]{*}{Transportation} & Density of Bus Network &\makecell[c]{The ratio of total length\\of bus routes to urban land area (km/km2)} \\
   \cmidrule{3-4}
    &  & Metro Network Characteristics &\makecell[c]{ Complex network indicators of the metro network,\\including network efficiency, average distance, and network density.} \\
   \cmidrule{2-4}  
   & \makecell[c]{Functional\\ Mixure} & POI   &\makecell[c]{POI entropy calculated based on Shannon entropy\\within a 500-meter radius of the station.} \\
   \cmidrule{2-4}  
   & \makecell[c]{Economic \\Vitality} & Nighttime Light Index &\makecell[c]{An effective representation of human activity. A larger value indicates\\ more frequent economic activity around the station \citep{jean2016combining}.} \\
\cmidrule{2-4}  
& Accessibility & Station Accessibility &\makecell[c]{Complex network indicators of the station,\\including node degree, closeness centrality, and node betweenness.} \\
\cmidrule{2-4}   
&\makecell[c]{ Station\\Function} &\makecell[c]{Origin-Destination Stations,\\Transfer Stations} &\makecell[c]{Whether it is an origin-destination station or a transfer station\\ (1 for yes, 0 for no). }\\
\cmidrule{2-4} 
& \makecell[c]{Transportation\\Facilities} & Number of Connecting Bus Stations & Number of bus stations within a 500-meter radius. \\
\cmidrule{2-4}  
& \makecell[c]{Geographic\\Location} & Distance to city center & Geographic distance to the city's economic center (km). \\
    \midrule
    \makecell[c]{Dynamic\\Covariates} & Weather & Temperature, Rainfall, AQI, etc. & Adverse weather conditions can reduce the number of residents' trips. \\
    \bottomrule
    \end{tabular}
  \label{tab1}
\end{table}

Given the synchronous correlation between dynamic feature and station passenger flow, such that weather factors impact passenger flow at current time intervals, a feature network is deemed useful for obtaining upcoming dynamic covariates. This can be portrayed as: 
\begin{subequations}
\begin{align}
&\tilde{\mathcal{D}}_{t}^{\boldsymbol{S}}=F_{D}^{\boldsymbol{S}}\left( \mathcal{D}_{t}^{\boldsymbol{S}} \right) \in \mathbb{R}^{D\times 1} \label{5a}\\
&\tilde{\mathcal{D}}_{t}^{\boldsymbol{G}}=F_{D}^{\boldsymbol{G}}\left( \mathcal{D}_{t}^{\boldsymbol{G}} \right) \in \mathbb{R}^{D\times 1}\label{5b}
\end{align}
\end{subequations}

Where $F_{D}^{\boldsymbol{S}}\left( \cdot \right) $ represents the feature network responsible for dynamic covariates, and $F_{D}^{\boldsymbol{G}}\left( \cdot \right) $ denotes the feature network with $F_{D}^{\boldsymbol{S}}\left( \cdot \right) $ set as the initial parameters.

\begin{figure}[ht]
	\centering
	\includegraphics[width=3.0cm]{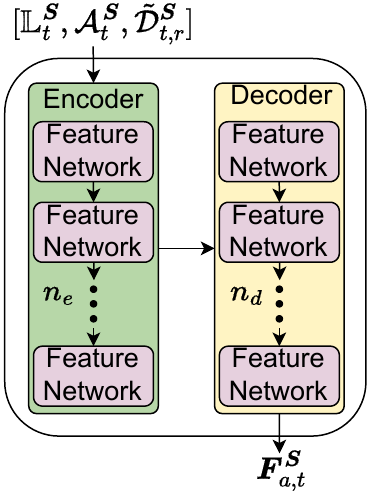}
	\caption{Encoder-Decoder module.}
	\label{lu5}
\end{figure}

Subsequently, the combination of passenger flow inputs, static covariates, and future dynamic covariates serves as inputs for the Encoder-Decoder module with the purpose of acquiring features. In the encoder, each input sequence is assigned a vector representation in a high-dimensional space. The resultant vector representation is deemed as a feature extraction of the input sequence. In the decoder, the initial input vector representation, alongside the formerly generated output sequence performs as input for delivering the subsequent output. In this study, the Encoder-Decoder module incorporates $n_e$ and $n_d$ concatenated feature networks, as demonstrated in Figure~\ref{lu5}. By amalgamating several feature networks, the model achieves parameter sharing, which reduces the number of parameters in the model leading to improved efficiency and generalization. What's more, it enables the construction of more complex nonlinear transformations that facilitate the model's adaptability to intricate input data.

To allocate dynamic covariates to each station, it is applied to the entire metro network. Subsequently, the historical passenger flow inputs and static covariates are merged and utilized as the inputs for the Encoder-Decoder module as follows: 
\begin{subequations}
\begin{align}
&\tilde{\mathcal{D}}_{t,r}^{\boldsymbol{S}}=Rep\left( \tilde{\mathcal{D}}_{t}^{\boldsymbol{S}} \right) \in \mathbb{R}^{S\times D}\label{6a}\\
&\tilde{\mathcal{D}}_{t,r}^{\boldsymbol{G}}=Rep\left( \tilde{\mathcal{D}}_{t}^{\boldsymbol{G}} \right) \in \mathbb{R}^{G\times D}\label{6b}\\
&\boldsymbol{F}_{a,t}^{\boldsymbol{S}}=F_{ED}^{\boldsymbol{S}}\left( con\left( \mathbb{L}_{t}^{\boldsymbol{S}},\mathcal{A}_{t}^{\boldsymbol{S}},\tilde{\mathcal{D}}_{t,r}^{\boldsymbol{S}} \right) \right) \in \mathbb{R}^{S\times emb}\label{6c}\\
&\boldsymbol{F}_{a,t}^{\boldsymbol{G}}=F_{ED}^{\boldsymbol{G}}\left( con\left( \mathbb{L}_{t}^{\boldsymbol{G}},\mathcal{A}_{t}^{\boldsymbol{G}},\tilde{\mathcal{D}}_{t,r}^{\boldsymbol{G}} \right) \right) \in \mathbb{R}^{G\times emb}\label{6d}
\end{align}
\label{eq6}
\end{subequations}

Where $Rep\left( \cdot \right)$ denotes the replication operation, $F_{ED}^{\boldsymbol{S}}\left( \cdot \right)$ refers to the representation of the Encoder-Decoder module during the pre-training phase, $F_{ED}^{\boldsymbol{G}}\left( \cdot \right)$ indicates the representation of the Encoder-Decoder module in the fine-tuning phase constraining $F_{ED}^{\boldsymbol{S}}\left( \cdot \right)$ as the initial parameters, $con\left( \cdot \right)$ symbolizes the concatenate operation and $emb$ represents the dimension of the feature embedding.
\subsubsection{Heterogeneous feature conversion}
Next we use transfer learning to transfer the learned station passenger flow feature embeddings from the source city to the target city. When performing transfer learning on the source and target city passenger flow data sets, the core is to measure the difference between the source domain and the target domain. We first use the station passenger flow data of the target city to construct the target data set, and then use the relevant station passenger flow data of the source city to form the source data set. Next, we calculate the distance between the station passenger flow feature embeddings in the source city and the target city to perform knowledge transfer between the source and target domains \citep{wang2018cross}. In this study, we calculate the weighted distance of paired stations passenger flow feature embeddings so that the trained feature network trained in the source city can guide the feature network of the target city to perform parameter optimization to improve generalization. Considering a target city station, $r^g$, optimally matched to a relevant station in the source city, $r^{g*}\in \boldsymbol{S}$ satisfies the following condition:
\begin{align}
P\left( \boldsymbol{x}_{T}^{g},\boldsymbol{x}_{T}^{g*} \right) \ge P\left( \boldsymbol{x}_{T}^{g},\boldsymbol{x}_{T}^{s} \right) ,\ \ \forall r^s\in \boldsymbol{S}
\label{eq7}
\end{align}

More so, the computation of the distance between station feature embeddings is achieved via the following expression:
\begin{align}
loss_{st}=\frac{1}{G}\sum_{r^g\in \boldsymbol{G}}{P\left( \boldsymbol{x}_{T}^{g},\boldsymbol{x}_{T}^{g*} \right) \cdot L\left( \boldsymbol{F}_{a,t}^{\boldsymbol{G}}\left( r^g \right) ,\boldsymbol{F}_{a,t}^{\boldsymbol{S}}\left( r^{g*} \right) \right)}
\label{eq8}
\end{align}

Where $\boldsymbol{F}_{a,t}^{\boldsymbol{G}}\left( r^g \right) \in \mathbb{R}^{1\times emb}$ and $\boldsymbol{F}_{a,t}^{\boldsymbol{S}}\left( r^{g*} \right) \in \mathbb{R}^{1\times emb}$ signify the feature embeddings for the target city station $r^g$ and the source city station $r^{g*}$ correspondingly; $L\left( \cdot \right) $ is the euclidean distance function. 
\subsubsection{Cross-city metro passenger flow prediction optimization}
In the preceding section, we minimized disparities in passenger flow characteristics between the source and target city. Subsequently, we employ the acquired feature embeddings to refine the prediction model for the target city. At first, we employ the historical passenger flow features of the target city to develop a basic prediction network, and obtain basic predicted value as follows:
\begin{align}
\boldsymbol{x}_{b}^{\boldsymbol{G}}=F^b\left( \mathbb{L}_{t}^{\boldsymbol{S}} \right) 
\label{eq9}
\end{align}

Here, $F^b\left( \cdot \right)$ represents the basic prediction network for the target city.

\begin{figure}[ht]
	\centering
	\includegraphics[width=7cm]{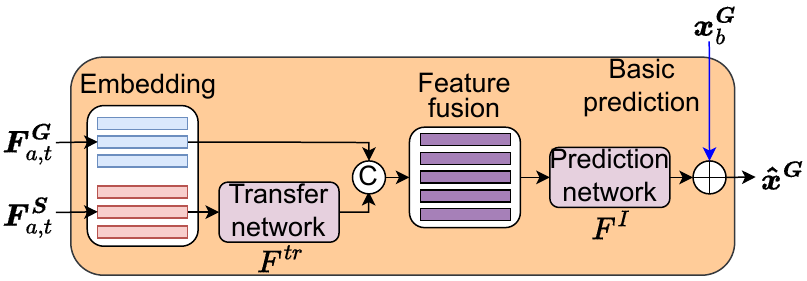}
	\caption{Feature fusion module.}
	\label{lu6}
\end{figure}

After minimizing differences in features between the source and target city, the learned feature embeddings are utilized to optimize the prediction model for the target city. Creation of a fusion module follows, which combines feature embeddings of both cities to obtain the initial and final prediction value as portrayed in Figure~\ref{lu6}. To begin with, we utilize the transformation network to align the dimensions of source and target city embedded features. We then proceed on to the concatenation of the embedded features vector and use it as input for the initial forecast network for the target city to derive the initial prediction value $I^{\boldsymbol{G}}$. Through the addition operation, we utilize the summation operation to fuse the basic prediction value $\boldsymbol{x}_{b}^{\boldsymbol{G}}$ and $I^{\boldsymbol{G}}$, thus deriving the final predicted passenger flow, $\boldsymbol{\hat{x}}^{\boldsymbol{G}}$. The basic prediction value, $\boldsymbol{x}_{b}^{\boldsymbol{G}}$ offers conformity to signal errors passing through $I^{\boldsymbol{G}}$ to avoid negative transfer from the source city. Additionally, it accelerates the model's convergence.
The mathematical process is outlined below:
\begin{subequations}
\begin{align}
&\boldsymbol{F}_{a,t}^{\boldsymbol{S'}}=F^{tr}\left( \boldsymbol{F}_{a,t}^{\boldsymbol{S}} \right) \label{10a}\\
&I^{\boldsymbol{G}}=F^I\left( con\left( \boldsymbol{F}_{a,t}^{\boldsymbol{S'}},\boldsymbol{F}_{a,t}^{\boldsymbol{G}} \right) \right) 
\label{10b}\\
&\boldsymbol{\hat{x}}^{\boldsymbol{G}}=add\left( \boldsymbol{x}_{b}^{\boldsymbol{G}},I^{\boldsymbol{G}} \right)\label{10c}
\end{align}
\label{eq10}
\end{subequations}

Where $\boldsymbol{F}_{a,t}^{\boldsymbol{S'}}$ denotes the transformed feature embedding for the source city; $F^{tr}$ and $F^I\left( \cdot \right)$ depict transformation and initial prediction network; $add\left( \cdot \right)$ signifies the addition operation.

The research's goal is to lower the prediction error of the passenger flow within the stations of the target city. To achieve this objective, the MAE (Mean Absolute Error) method replaces the conventional MSE (Mean Squared Error), thereby reducing the model's vulnerability to atypical values \citep{han2023capacity}:
\begin{align}
loss_r=\frac{1}{G}\sum_{\boldsymbol{G}}{\left| \boldsymbol{\hat{x}}^{\boldsymbol{G}}-\boldsymbol{x}^{\boldsymbol{G}} \right|}
\label{eq11}
\end{align}

Finally, The loss function of the model blends two components, specifically the prediction error and feature embedding loss. Consequently, the general loss, $loss_f$, is defined as:
\begin{align}
loss_f=\left( 1-w \right) \cdot loss_r+w\cdot loss_{st}
\label{eq12}
\end{align}

Where $w$ signifies the balance coefficient, determining the equilibrium between the feature embedding variation and the prediction error.
\subsubsection{Training Process}
The overall training process of METcross consists of two steps: source city pre-training and optimization of metro passenger flow prediction for the target city. During the pre-training phase, the feature extraction module $F_{D}^{\boldsymbol{S}}\left( \cdot \right)$, Encoder-Decoder module $F_{ED}^{\boldsymbol{S}}\left( \cdot \right)$ and prediction network $Pre_{\boldsymbol{S}}\left( \cdot \right)$ are trained.

During the fine-tuning phase, the pre-training model is utilized to derive the feature embeddings $\boldsymbol{F}_{a,t}^{\boldsymbol{S}}$ and $\boldsymbol{F}_{a,t}^{\boldsymbol{G}}$ for both the source and target cities, correspondingly. Afterward, the feature embedding loss $loss_{st}$ is computed. Subsequently, we acquire the initial prediction flow $I^{\boldsymbol{G}}$ utilizing the embedding features of both the source and target city. Furthermore, we construct the basic prediction model via employing the historical passenger flow input features of the target city to determine the basic prediction values $\boldsymbol{x}_{b}^{\boldsymbol{G}}$. The final prediction value $\boldsymbol{\hat{x}}^{\boldsymbol{G}}$ is attained by combining $\boldsymbol{x}_{b}^{\boldsymbol{G}}$ and $I^{\boldsymbol{G}}$ through residual connections. To compare the actual passenger flow $\boldsymbol{x}^{\boldsymbol{G}}$ with the predictions, we calculate the prediction error $loss_r$. Finally, the balanced feature embedding loss and prediction error loss are fused. We summarize the algorithmic steps for pre-training in the source city and optimizing the metro passenger flow prediction for the target city in Algorithm~\ref{algorithm1} and Algorithm~\ref{algorithm2}.
\begin{algorithm}[ht]
	\caption{Pre-training Process in the Source City }\label{algorithm1}
	\LinesNumbered 
	\KwIn{Source city passenger flow input matrix $\mathbb{L}_{t}^{\boldsymbol{S}}$; Source city static and dynamic covariate feature matrix $\mathcal{A}_{t}^{\boldsymbol{S}}$, $\mathcal{D}_{t}^{\boldsymbol{S}}$; Random initial parameters $\theta$}
	\KwOut{Source city model parameters $\theta ^{\boldsymbol{S}}$, including $F_{D}^{\boldsymbol{S}}$, $F_{ED}^{\boldsymbol{S}}$ and $Pre_{\boldsymbol{S}}$}
	\While{epoch $\leq$ Epochs;}
        {\For  {t $\in$ T}
        {Get $\boldsymbol{F}_{a,t}^{\boldsymbol{S}}$ \ by \ Eq ~(\ref{6c});\\
        Get $\boldsymbol{\hat{x}}^{\boldsymbol{S}}=Pre_{\boldsymbol{S}}\left(\boldsymbol{F}_{a,t}^{\boldsymbol{S}}\right)$;\\
        $loss=\frac{1}{S}\sum_{\boldsymbol{S}}{\left| \boldsymbol{\hat{x}}^{\boldsymbol{S}}-\boldsymbol{x}^{\boldsymbol{S}} \right|}$;\\
        $\theta$  $\gets$ $loss$;
        }}  
    $\theta^{\boldsymbol{S}}$ $\gets$ $\theta$; \\
    \Return{$\theta ^{\boldsymbol{S}}$}
\end{algorithm}

\begin{algorithm}[ht]
	\caption{Optimal Prediction of Metro Passenger Flow in Target City}\label{algorithm2}
	\LinesNumbered 
	\KwIn{Source and target city passenger flow input matrix $\mathbb{L}_{t}^{\boldsymbol{S}}$, $\mathbb{L}_{t}^{\boldsymbol{G}}$; Static covariate feature matrix $\mathcal{A}_{t}^{\boldsymbol{S}}$, $\mathcal{A}_{t}^{\boldsymbol{G}}$; Dynamic covariate feature matrix $\mathcal{D}_{t}^{\boldsymbol{S}}$, $\mathcal{D}_{t}^{\boldsymbol{G}}$; Model parameters for the source city $\theta ^{\boldsymbol{S}}$; Initialize $\theta ^{\boldsymbol{S}}$ as the initial parameter $\theta$}
	\KwOut{Target city model parameters $\theta ^{\boldsymbol{G}}$, including $F_{D}^{\boldsymbol{G}}$, $F_{ED}^{\boldsymbol{G}}$, $F^b$, $F^{tr}$ and $F^f$}
	\While{epoch $\leq$ Epochs;}
        {\For  {t $\in$ T}
        {Get $\boldsymbol{F}_{a,t}^{\boldsymbol{S}}$ \ and \ $\boldsymbol{F}_{a,t}^{\boldsymbol{G}}$ \ by \ Eqs ~(\ref{6c}) and ~(\ref{6d}) ;\\
        Get $loss_{st}$ \ by \ Eq ~(\ref{eq8});\\
        Get $\boldsymbol{x}_{b}^{\boldsymbol{G}}$ \ by \ Eq ~(\ref{eq9}) ;\\
        Get  $\boldsymbol{\hat{x}}^{\boldsymbol{G}}$ \ by \ Eq ~(\ref{10c}) ;\\
        Get $loss_{r}$ by \ Eq ~(\ref{eq11}) ;\\
        Get $loss_{f}$ by \ Eq ~(\ref{eq12}) ;\\
        $\theta$  $\gets$ $loss_{f}$;
        }}
    $\theta ^{\boldsymbol{G}}$ $\gets$ $\theta$;\\
    \Return{$\theta ^{\boldsymbol{G}}$}
\end{algorithm}

\section{Case Study}
\subsection{Experiment Settings}
\subsubsection{Data Description}
\begin{figure}[ht]
	\centering
	\includegraphics[width=11cm]{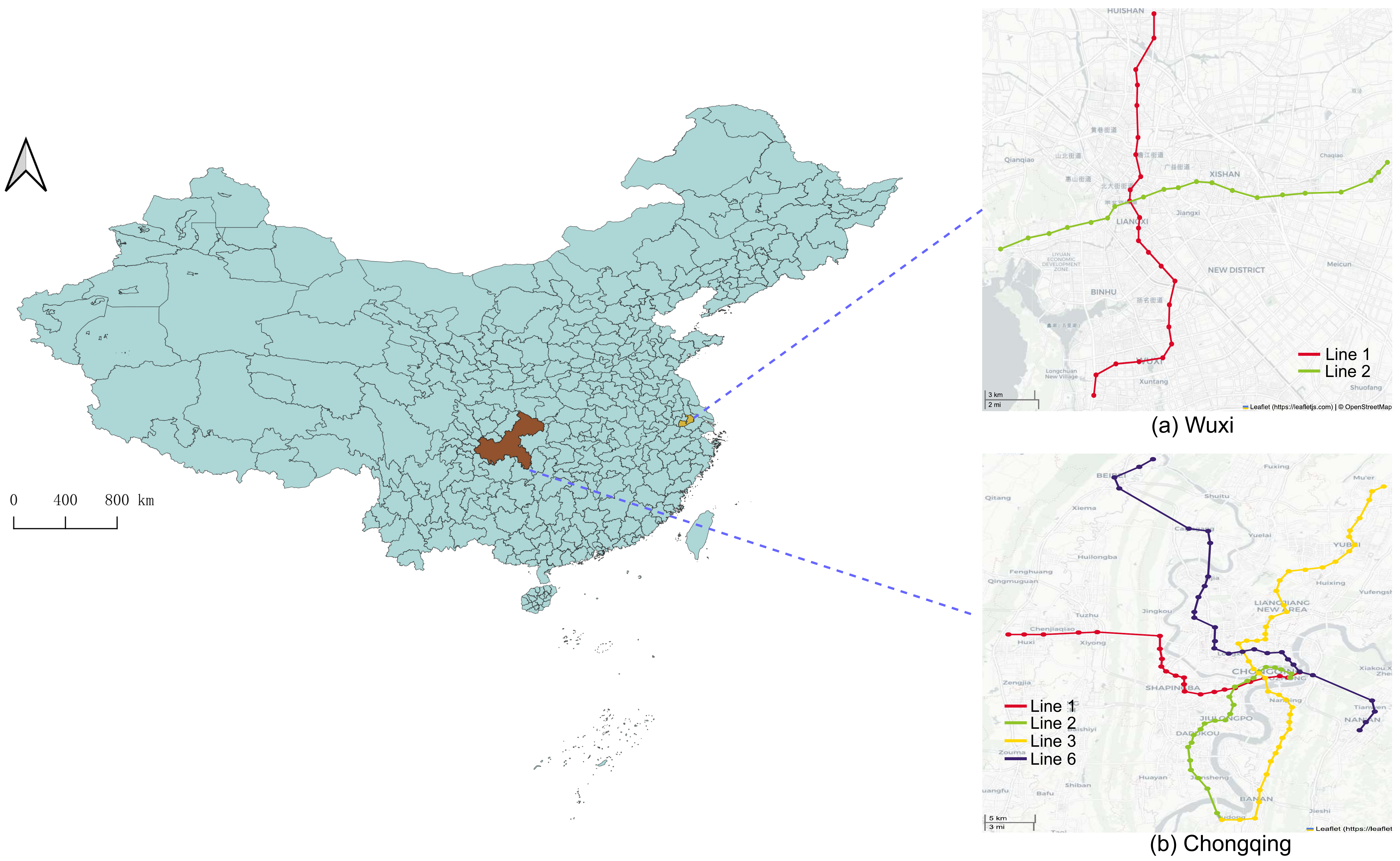}
	\caption{Study area.}
	\label{lu7}
\end{figure}

As demonstrated in Figure~\ref{lu7}, we adopt the metro systems of two Chinese cities as exemplary cases in this research. Wuxi, featuring two lines and 44 stations, and Chongqing, with four lines and 115 stations, are examined. The inflow passenger volume is computed by aggregating AFC (Automatic Fare Collection) data at a time granularity of 10 minutes, spanning from March 1st through March 31st, 2017. To test the accuracy of our model, we utilize the last 5 days as test data. Furthermore, we set the length of training data to be 20, 7, and 3 days, respectively. The study collected external information from various sources. Population and public transportation network data were sourced from the China Statistical Yearbook, while Gaode Maps provided POI data, bus route numbers, and geographical locations. The nighttime light index was retrieved from Harvard Dataverse \citep{DVN/GIYGJU_2021}. The weather data was gathered from the National Tibetan Plateau Data Center (\url{https://data.tpdc.ac.cn/home}), as Table~\ref{tab2} shows.

\begin{table}[ht]
  \centering
  \scriptsize
  \caption{Examples of weather data.}
    \begin{tabular}{ccccccccccccc}
    \toprule
    Date&	Hour&\makecell[c]{Temperature\\(℃)}&\makecell[c]{Humidity\\(\%)}&	\makecell[c]{Rain\\(mm/h)}&\makecell[c]{Wind\\(m/s)}&	AQI&	PM2.5&	PM10&	$SO_2$	&$NO_2$&	$O_3$&CO\\
    \midrule
    20170301 & 6     & 8.96 & 89.83 & 0     & 1.34 & 80    & 59    & 96    & 17    & 63    & 11    & 1.01 \\
    20170301 & 7     & 9.33 & 86.80 & 0     & 1.06 & 81    & 60    & 94    & 17    & 66    & 8     & 1.09 \\
    20170301 & 8     & 9.51 & 84.75 & 0     & 1.36 & 84    & 62    & 96    & 18    & 66    & 7     & 1.3 \\
    20170301 & 9     & 14.27 & 64.91 & 0     & 1.35 & 88    & 65    & 109   & 20    & 69    & 7     & 1.44 \\
    20170301 & 10    & 14.05 & 65.82 & 0     & 1.75 & 93    & 69    & 118   & 25    & 72    & 8     & 1.36 \\
    20170301 & 11    & 14.31 & 66.42 & 0     & 2.17 & 95    & 71    & 127   & 35    & 79    & 16    & 1.31 \\
    20170301 & 12    & 17.89 & 51.07 & 0     & 2.39 & 99    & 74    & 135   & 37    & 77    & 31    & 1.14 \\
    20170301 & 13    & 18.54 & 46.93 & 0.09 & 1.52 & 95    & 71    & 121   & 32    & 60    & 54    & 0.93 \\
    \bottomrule
    \end{tabular}
  \label{tab2}
\end{table}

\subsubsection{Programming Details}
Our computer system is an i5-12490F CPU equipped with an RTX2060 GPU that contains 6GB of memory. We established the subsequent hyperparameter configurations: the model consists of one hidden layer with 128 hidden neurons, one feature network block in both the Encoder and Decoder. Additionally, a batch size of 256 and a dropout rate of 0 were implemented. The learning rate was set to 0.001, with a benchmark of 100 epochs, and Adam optimization algorithm. 

\subsection{Model Evaluation}
In this research, we chose MAE and RMSE (Root Mean Squared Error) as the evaluation metrics to gauge the model's effectiveness. To address multivariate prediction matters, we calculated the average of the prediction indicators across multiple stations, as follows:
\begin{subequations}
\begin{align}
&MAE=\frac{1}{G}\sum_{g=1}^G{MAE_g},\ MAE_g=\frac{1}{T'}\sum_{t=1}^{T'}{\left| \hat{x}_{t}^{g}-x_{t}^{g} \right|}\\
&RMSE=\frac{1}{G}\sum_{g=1}^G{RMSE_g},\ RMSE_g=\sqrt{\frac{1}{T'}\sum_{t=1}^{T'}{\left( \hat{x}_{t}^{g}-x_{t}^{g} \right) ^2}}
\end{align}
\end{subequations} 

where $T'$ is the number of samples in the test set.
\subsection{Model Comparison}
We use the MLP and LSTM  as baseline networks. Baselines were constructed from five presentations: NF, DF, NF, PF, and FT. The names of each model are listed in Table~\ref{tab3}.
\begin{table}[ht]
  \scriptsize
  \setlength{\tabcolsep}{1.5mm}
  \centering
  \caption{Various baseline models.}
    \begin{tabular}{cccccccccccccc}
    \toprule
    \multirow{2}[4]{*}{Predictor} & \multirow{2}[4]{*}{NF} &\multicolumn{3}{c}{DF} &\multicolumn{3}{c}{FF} &\multicolumn{3}{c}{PF} &\multirow{2}[4]{*}{FT-P} &\multirow{2}[4]{*}{FT-F} &\multirow{2}[4]{*}{ METcross} \\
\cmidrule{3-11}    
   & & AJ    & We    & Si    & AJ    & We    & Si    & AJ    & We    & Si    &  &  &  \\
    \midrule
    MLP   & MLP   & \makecell[c]{MLDF\\\_AJ} &  \makecell[c]{MLDF\\\_We} &  \makecell[c]{MLDF\\\_Si} &  \makecell[c]{MLFF\\\_AJ}&  \makecell[c]{MLFF\\\_We} & \makecell[c]{MLFF\\\_Si} &  \makecell[c]{MLPF\\\_AJ} &  \makecell[c]{MLPF\\\_We} & \makecell[c]{ MLPF\\\_Si} &  ML\_FT-P &  ML\_FT-F & ML\_Cross \\
    \midrule
    LSTM  & LSTM  & \makecell[c]{ LSDF\\\_AJ} &  \makecell[c]{LSDF\\\_We }&  \makecell[c]{LSDF\\\_Si} & \makecell[c]{LSFF\\\_AJ} & \makecell[c]{LSFF\\\_We} &  \makecell[c]{LSFF\\\_Si} &  \makecell[c]{LSPF\\\_AJ }&  \makecell[c]{LSPF\\\_We} &  \makecell[c]{LSPF\\\_Si} & LS\_FT-P & LS\_FT-F & LS\_Cross \\
    \bottomrule
    \end{tabular}
  \label{tab3}
\end{table}
\subsection{Analysis of Prediction Results}
The models were trained with the designated training data and employed for predicting results, as shown in Tables~\ref{tab4} and~\ref{tab5}. Each row labels the superior results in bold black, with the second-best shown in underline. The percentage reduction in error by the METcross framework compared to NF is labeled "Boost". Our observations are as follows: 
\begin{table}[ht]
  \scriptsize
  \centering
  \setlength{\tabcolsep}{0.9mm}
  \caption{Prediction metrics based on MLP as the base network.}
    \begin{tabular}{ccccccccccccccccc}
   \toprule
   Target&\makecell[c]{Train\\data} & Metric & MLP   & \makecell[c]{MLDF\\\_AJ} &  \makecell[c]{MLDF\\\_We} &  \makecell[c]{MLDF\\\_Si} &  \makecell[c]{MLFF\\\_AJ}&  \makecell[c]{MLFF\\\_We} & \makecell[c]{MLFF\\\_Si} &  \makecell[c]{MLPF\\\_AJ} &  \makecell[c]{MLPF\\\_We} & \makecell[c]{ MLPF\\\_Si} & \makecell[c]{ ML\\\_FT-P }&  \makecell[c]{ML\\\_FT-F} & \makecell[c]{ML\\\_Cross} & Boost(\%) \\
    \midrule
    {\multirow{6}[12]{*}{WX}} &{\multirow{2}[4]{*}{\makecell[c]{25\\days}}} & MAE   & 10.048 & 11.361 & 10.901 & 16.006 & 9.687 & 9.597 & \underline{9.267} & 9.824 & 9.822 & 9.377 & 9.982 & 9.892 & \textbf{8.397} & 16.431 \\
\cmidrule{3-17}          &       & RMSE  & 14.285 & 15.879 & 15.287 & 22.401 & 13.783 & 13.641 & 13.169 & 13.955 & 13.967 & \underline{13.121} & 14.22 & 14.068 & \textbf{11.663} & 18.355 \\
\cmidrule{2-17}          & {\multirow{2}[4]{*}{\makecell[c]{7\\days}}} & MAE   & 10.189 & 11.388 & 10.885 & 17.834 & 9.824 &\underline{ 9.748} & 10.045 & 9.918 & 9.949 & 10.369 & 10.034 & 9.953 & \textbf{8.872} & 12.926 \\
\cmidrule{3-17}          &       & RMSE  & 14.488 & 15.984 & 15.319 & 24.729 & \underline{13.766} & 13.807 & 14.185 & 14.053 & 14.105 & 13.822 & 14.273 & 14.212 & \textbf{12.428} & 14.219 \\
\cmidrule{2-17}          & {\multirow{2}[4]{*}{\makecell[c]{3\\days}}} & MAE   & 10.467 & 11.652 & 11.174 & 16.772 &\underline{ 10.033} & 10.058 & 10.208 & 10.057 & 10.035 & \textbf{9.966} & 10.13 & 10.086 & 10.117 & 3.344 \\
\cmidrule{3-17}          &       & RMSE  & 14.868 & 16.288 & 15.64 & 23.233 & 14.283 & 14.344 & 14.302 & 14.229 &\underline{14.209} & \textbf{14.082} & 14.457 & 14.389 & 14.22 & 4.358 \\
    \midrule
    {\multirow{6}[12]{*}{CQ}} & {\multirow{2}[4]{*}{\makecell[c]{25\\days}}} & MAE   & 20.522 & 21.907 & 20.945 & 33.871 & 18.411 & 18.283 &\underline{18.155} & 18.945 & 18.882 & 18.78 & 20.311 & 19.895 & \textbf{15.935} & 22.352 \\
\cmidrule{3-17}          &       & RMSE  & 31.038 & 32.04 & 30.96 & 48.686 & 27.513 & 27.367 & \underline{27.173} & 28.548 & 28.489 & 28.552 & 30.667 & 30.13 & \textbf{23.27} & 25.027 \\
\cmidrule{2-17}          & {\multirow{2}[4]{*}{\makecell[c]{7\\days}}} & MAE   & 21.071 & 22.343 & 21.405 & 33.67 & 19.625 &\underline{18.894} & 19.073 & 19.139 & 19.17 & 20.264 & 20.494 & 20.021 & \textbf{16.678} & 20.849 \\
\cmidrule{3-17}          &       & RMSE  & 31.977 & 32.71 & 31.695 & 50.299 & 30.247 & \underline{28.397} & 29.033 & 28.552 & 28.561 & 28.775 & 31.012 & 30.004 & \textbf{24.465} & 23.492 \\
\cmidrule{2-17}          & {\multirow{2}[4]{*}{\makecell[c]{3\\days}}} & MAE   & 22.142 & 22.83 & 21.907 & 34.114 & 19.311 & 19.347 & \underline{19.109} & 19.605 & 19.56 & 19.149 & 20.995 & 20.48 & \textbf{18.718} & 15.464 \\
\cmidrule{3-17}          &       & RMSE  & 34.28 & 33.807 & 32.663 & 51.687 & 29.119 & 29.219 & \underline{28.67} & 29.472 & 29.503 & 28.833 & 32.135 & 31.144 & \textbf{27.684} & 19.242 \\
   \bottomrule
    \end{tabular}
  \label{tab4}
\end{table}
\begin{table}[ht]
 \scriptsize
  \centering
  \setlength{\tabcolsep}{0.9mm}
  \caption{Prediction metrics based on LSTM as the base network.}
    \begin{tabular}{ccccccccccccccccc}
    \toprule
    Target&\makecell[c]{Train\\data} & Metric & LSTM  & \makecell[c]{ LSDF\\\_AJ} &  \makecell[c]{LSDF\\\_We }&  \makecell[c]{LSDF\\\_Si} & \makecell[c]{LSFF\\\_AJ} & \makecell[c]{LSFF\\\_We} &  \makecell[c]{LSFF\\\_Si} &  \makecell[c]{LSPF\\\_AJ }&  \makecell[c]{LSPF\\\_We} &  \makecell[c]{LSPF\\\_Si} & \makecell[c]{ LS\\\_FT-P }&  \makecell[c]{LS\\\_FT-F} & \makecell[c]{LS\\\_Cross} & Boost(\%) \\
    \midrule
    {\multirow{6}[12]{*}{WX}} & {\multirow{2}[4]{*}{\makecell[c]{25\\days}}} & MAE   & 9.063 & 9.785 & 9.634 & 11.951 & 8.351 & 8.551 & \underline{8.148} & 8.635 & 8.564 & 8.499 & 8.556 & 8.469 & \textbf{7.58} & 16.363 \\
\cmidrule{3-17}          &       & RMSE  & 12.831 & 13.725 & 13.626 & 17.292 & 11.768 & 12.043 & \underline{11.406} & 12.255 & 12.115 & 11.845 & 11.927 & 11.698 & \textbf{10.548} & 17.793 \\
\cmidrule{2-17}          & {\multirow{2}[4]{*}{\makecell[c]{7\\days}}} & MAE   & 9.714 & 10.571 & 10.117 & 12.64 & 9.426 & 9.241 & \underline{8.789} & 9.44  & 9.286 & 9.382 & 9.093 & 8.836 & \textbf{8.106} & 16.553 \\
\cmidrule{3-17}          &       & RMSE  & 13.638 & 14.724 & 14.162 & 17.879 & 13.169 & 13.03 & 12.424 & 13.325 & 13.042 & 13.216 & 12.809 & \underline{12.389} & \textbf{11.315} & 17.033 \\
\cmidrule{2-17}          & {\multirow{2}[4]{*}{\makecell[c]{3\\days}}} & MAE   & 9.865 & 11.106 & 10.708 & 13.211 & 9.586 & 9.572 & 9.919 & 9.581 & 9.564 & 9.635 & \underline{9.385} & 9.444 & \textbf{8.903} & 9.752 \\
\cmidrule{3-17}          &       & RMSE  & 13.995 & 15.441 & 14.863 & 18.472 & 13.59 & 13.475 & 13.629 & 13.428 & 13.43 & 13.497 & \underline{13.243} & 13.284 & \textbf{12.716} & 9.139 \\
    \midrule
    {\multirow{6}[12]{*}{CQ}} & {\multirow{2}[4]{*}{\makecell[c]{25\\days}}} & MAE   & 18.405 & 19.613 & 18.969 & 26.126 & 16.917 & 16.953 & \underline{16.596} & 17.156 & 17.151 & 17.615 & 17.807 & 17.769 & \textbf{14.349} & 22.037 \\
\cmidrule{3-17}          &       & RMSE  & 27.26 & 27.644 & 27.047 & 38.227 & 24.543 & 24.604 & \underline{23.661} & 24.527 & 24.733 & 26.031 & 26.028 & 25.846 & \textbf{20.124} & 26.178 \\
\cmidrule{2-17}          & {\multirow{2}[4]{*}{\makecell[c]{7\\days}}} & MAE   & 20.106 & 21.343 & 20.763 & 29.074 & 18.143 & 18.539 & 18.123 & 18.449 & 18.582 & 18.501 & 18.147 & \underline{17.794} & \textbf{16.295} & 18.955 \\
\cmidrule{3-17}          &       & RMSE  & 30.088 & 30.53 & 30.152 & 43.054 & 26.728 & 26.962 & 26.909 & 27.314 & 27.857 & 27.272 & 26.539 & \underline{25.735} & \textbf{23.912} & 20.526 \\
\cmidrule{2-17}          & {\multirow{2}[4]{*}{\makecell[c]{3\\days}}} & MAE   & 20.455 & 21.87 & 21.213 & 29.995 & 18.736 & 18.84 & 18.632 & 18.874 & 18.927 & 18.878 & 18.72 & \underline{18.494} & \textbf{17.784} & 13.058 \\
\cmidrule{3-17}          &       & RMSE  & 30.675 & 31.795 & 31.177 & 44.333 & 27.716 & 28.123 & 27.916 & 28.107 & 28.275 & 28.182 & 27.793 & \underline{27.158} & \textbf{26.447} & 13.783 \\
    \bottomrule
    \end{tabular}%
  \label{tab5}%
\end{table}%
\begin{itemize}
\item The METcross framework outperforms all models, with the maximum reduction being 22.35\% in MAE and 26.18\% in RMSE. 
\item A decrease in training length results in an increase in testing error. When training data is less, it finds it challenging to comprehend intricate patterns, affecting its accuracy. 
\item The DF model yields more significant prediction errors compared to the NF model, implying that direct input data fusion results in increased prediction errors. 
\item Fine-tuning of prediction and feature levels  from the source and target cities together effectively reduces prediction errors.
\end{itemize}

\begin{figure}[ht]
	\centering
	\includegraphics[width=10cm]{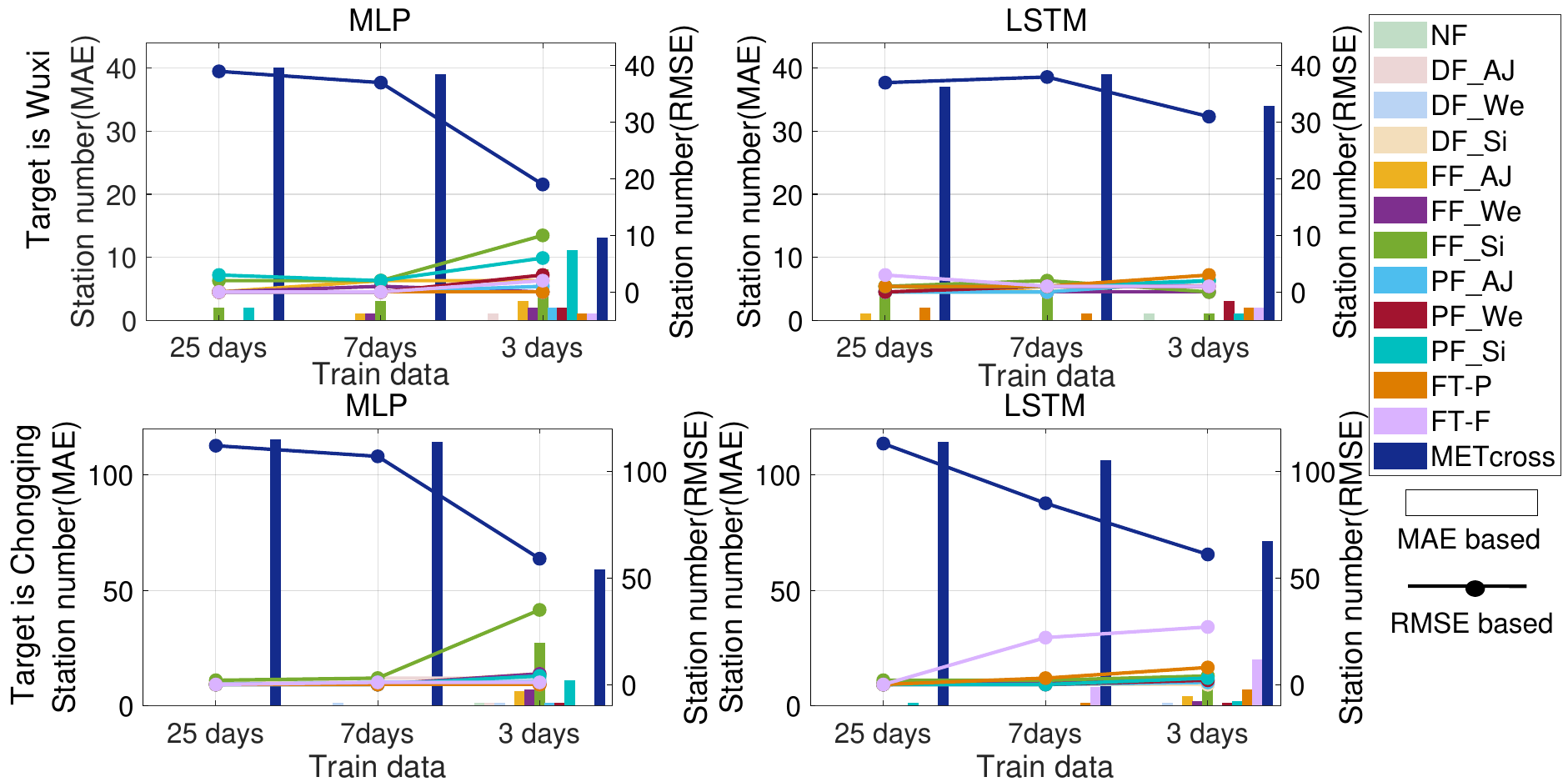}
	\caption{Number of best predicted stations for each method.}
	\label{lu8}
\end{figure}

Additionally, we examine the station predictive indicators of various models. We determine the ideal number of indicators for each model by comparing the stations' predictive indicators, which is displayed in Figure ~\ref{lu8}. METcross demonstrates the best performance in predicting the majority of stations' indicators. The optimal indicator count results vary little between training data intervals of 25 and 7 days. Furthermore, even with only 3 days of training data, METcross can still outperform other models in predicting most stations' indicators.

\begin{figure}[ht]
	\centering
	\includegraphics[width=11cm]{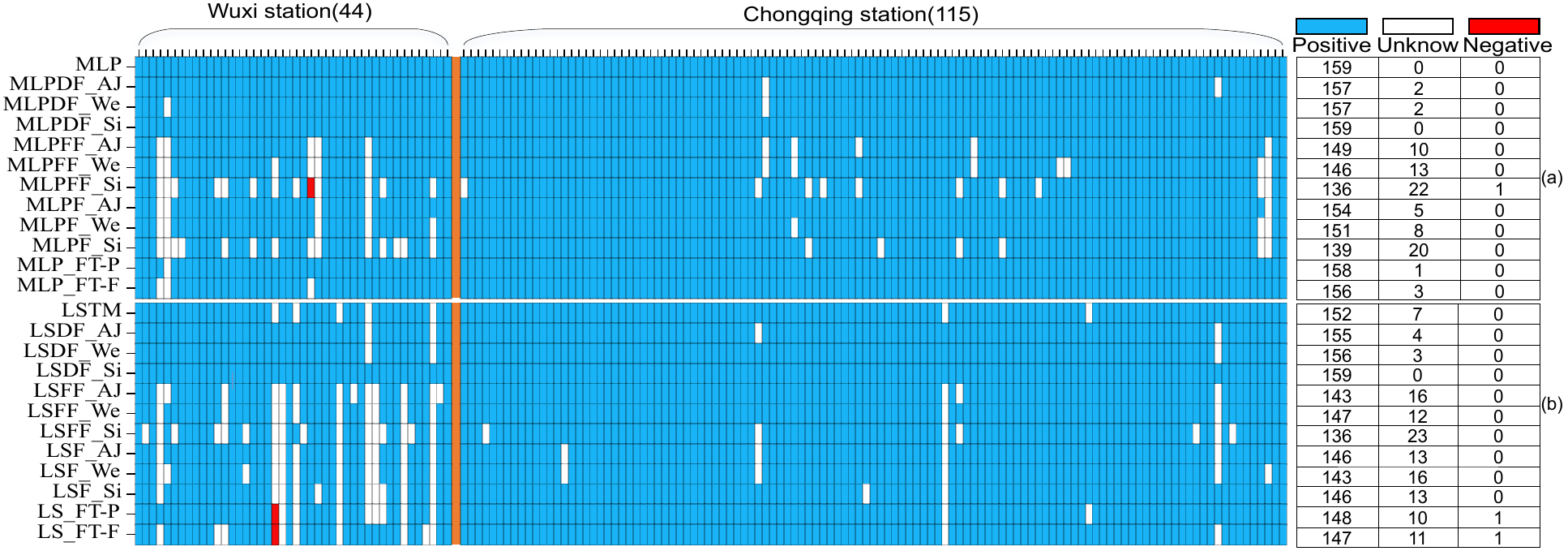}
	\caption{DM test results for each method. (a) MLP is the base method, with time lengths of 25 days; (b) LSTM is the base method, with time lengths of 25 days.
     }
\label{lu9}
\end{figure}

Next, we employ the Diebold-Mariano test \citep{diebold2002comparing} to statistically assess the prediction error of both the METcross framework and baseline models at each station. Figure~\ref{lu9} depicts the DM test results for both the METcross framework and baseline models at each station. DM values below 0 and p-values under 0.05 prompt the blue cell displaying significant predictive accuracy for the METcross framework. Conversely, DM values above 0 and p-values under 0.05 are linked to the red cell indicating better performance by the baseline method. P-values exceeding 0.05 are presented in a white cell, indicating uncertainty. The METcross framework exhibits higher predictive accuracy compared to baseline models at several stations.
\subsection{Basic Framework Analysis}
The prediction errors highlight the significant influence of the fusion, transformation, and fine-tuning methods. Consequently, analyzing the prediction errors associated with these methods is crucial in constructing suitable cross-city STPF prediction models.

\subsubsection{Fusion Methods}
\begin{figure}[ht]
	\centering
	\includegraphics[width=11cm]{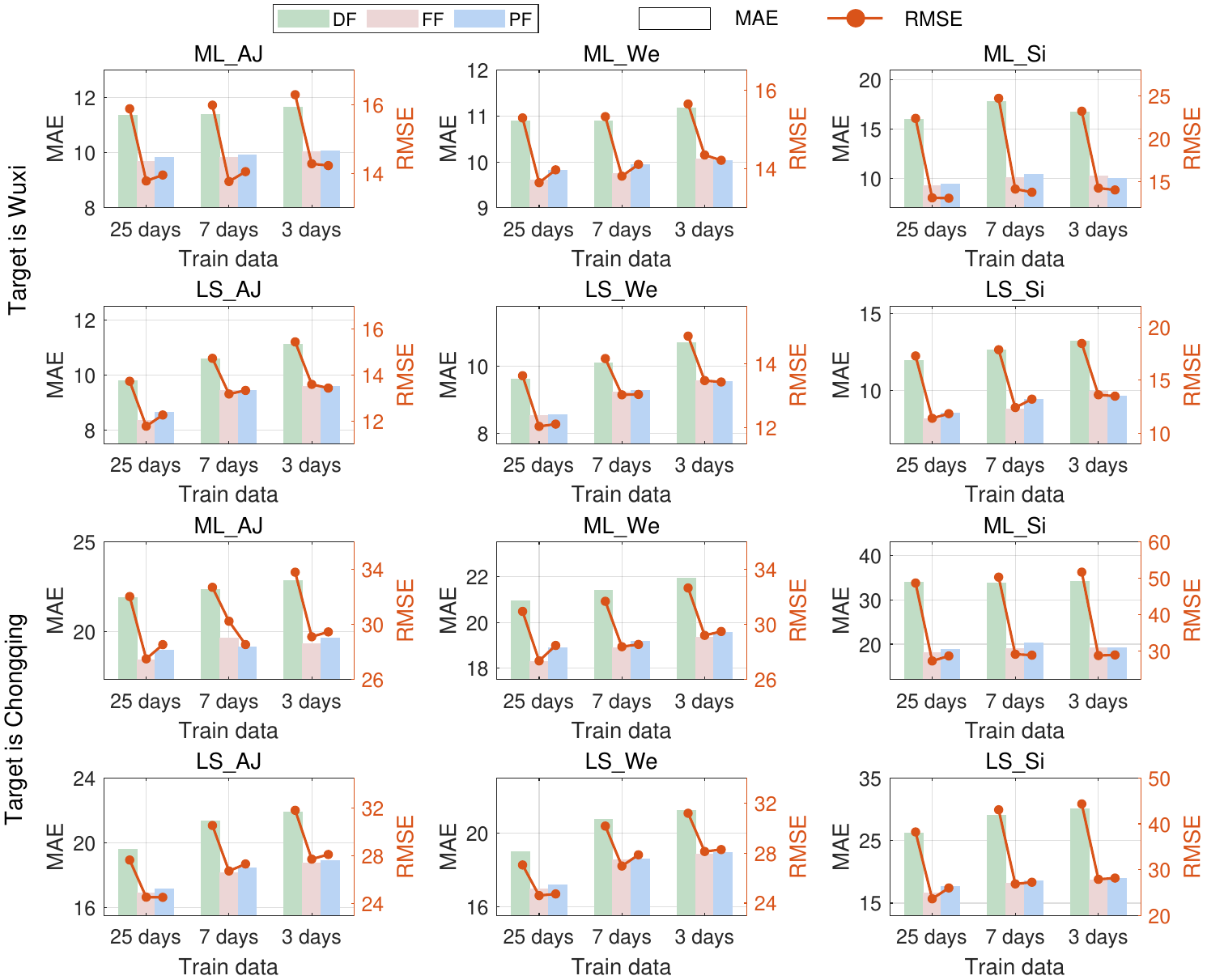}
	\caption{Prediction errors for different fusion methods.}
	\label{lu10}
\end{figure}

The prediction indicators were analyzed regarding input data fusion, feature fusion, and prediction result fusion, as illustrated in Figure~\ref{lu10}. Performance variations among various fusion techniques are apparent with changing training data length. Feature fusion, in most instances, yields superior results compared to connecting input data fusion directly. The selection of feature fusion and prediction result fusion techniques is related to the magnitude of metro networks. Utilizing the prediction result fusion approach is advisable in cases where the source city's training data is more extensive than the the target city.
\subsubsection{Transformation Methods}
\begin{figure}[ht]
	\centering
	\includegraphics[width=11cm]{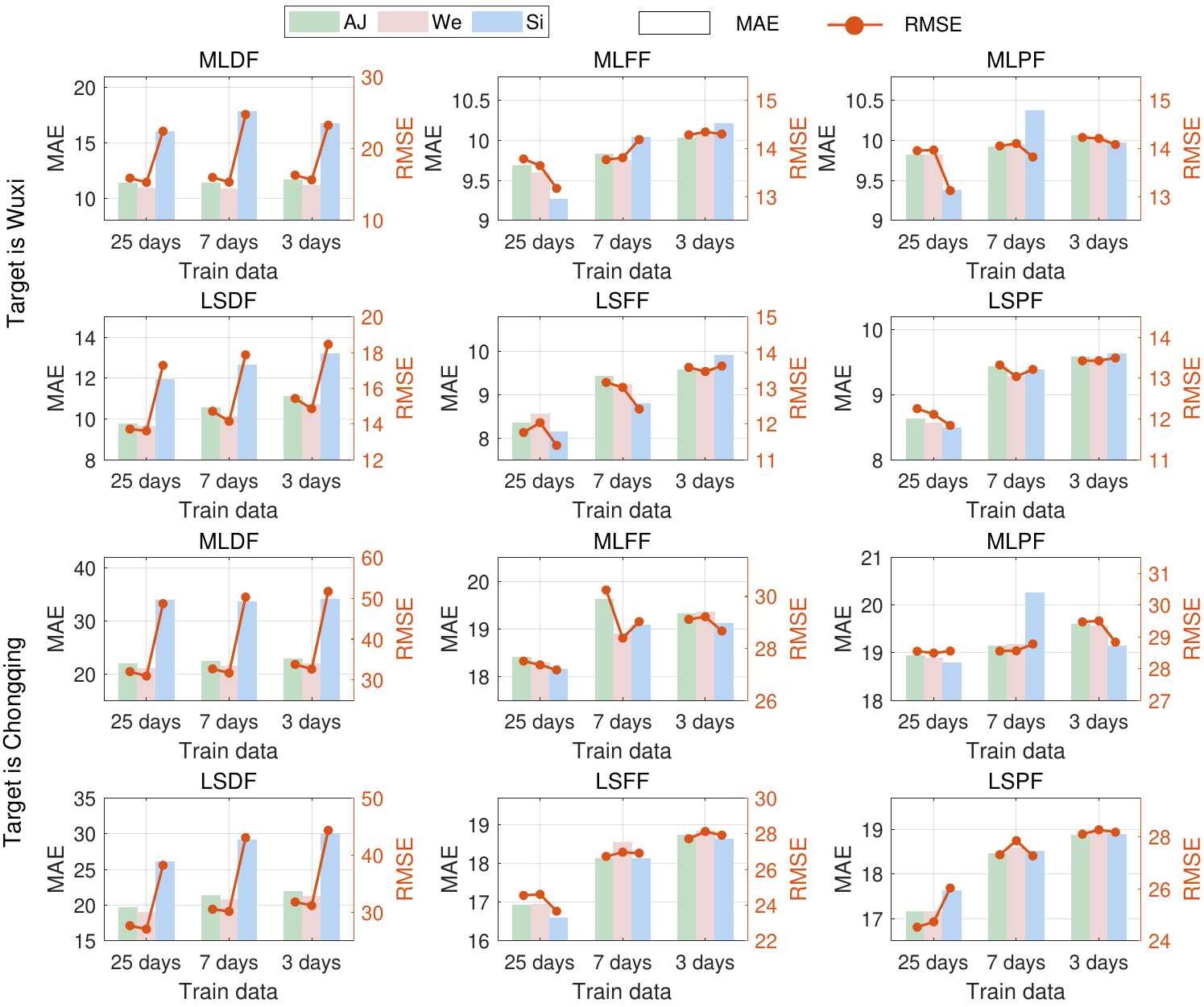}
	\caption{Prediction errors for different transformation methods.}
	\label{lu11}
\end{figure}

The impact of adjacency, weight, and similarity matrices on prediction indicators was analyzed, as indicated in Figure~\ref{lu11}. Notably, diverse transformation methods have a noticeable impact on prediction indicators. Both the adjacency and weight matrices solely rely on passenger flow information between matched stations. However, the weight matrix, by considering station similarity, generally yields higher performance than the adjacency matrix. Modeling risks escalate when employing the similarity matrix, which utilizes passenger flow information from all source city stations for each target city station. Using the weight matrix as the transformation approach usually guarantees accurate predictions while preserving a certain level of robustness.
\subsubsection{Fine-tuning Methods}
\begin{figure}[ht]
	\centering
	\includegraphics[width=8cm]{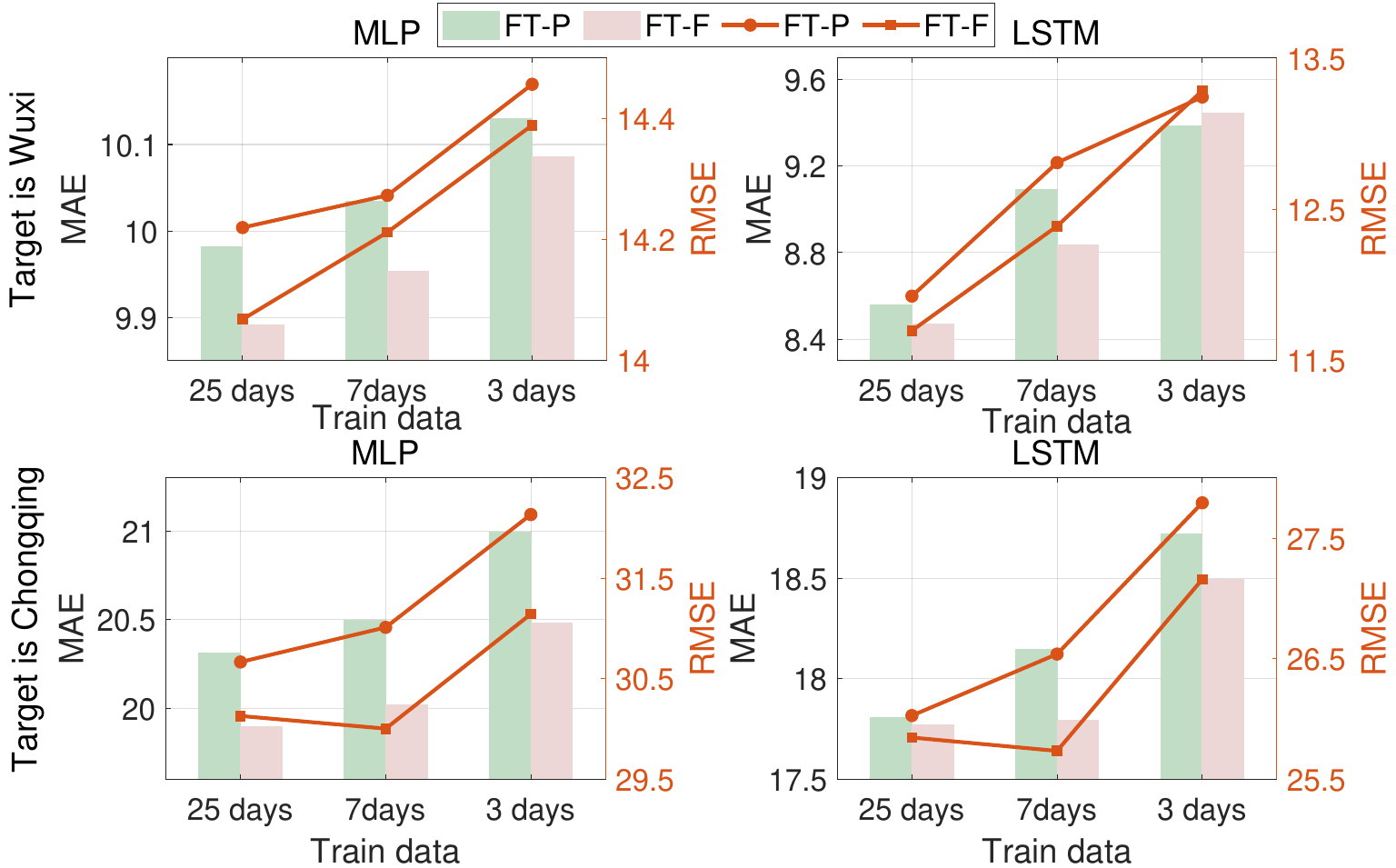}
	\caption{Prediction errors for different fine-tuning methods.}
	\label{lu12}
\end{figure}

Fine-tuning is a useful technique in reducing training time, unlike fusion methods. Fine-tuning involves utilizing the parameters of the pre-trained model as the initial parameters on the target city. We explored the impact of distinct pre-training models on prediction results from a transfer learning standpoint, as depicted in Figure~\ref{lu12}. Incorporating a feature network into the pre-trained model trained on the source city drastically minimizes prediction errors. It allows for the transfer of source city knowledge to the target city, enabling the accurate training of a passenger flow prediction model using data from the source city. 
\subsection{METcross Framework Analysis}
The METcross framework comprises several hyperparameters, which includes the loss function balance coefficient, feature embedding dimension, and feature network number. Determining suitable hyperparameters serves as a reference point for engineering practice.

\subsubsection{Influence of Balance Coefficient}
\begin{figure}[ht]
	\centering
	\includegraphics[width=11cm]{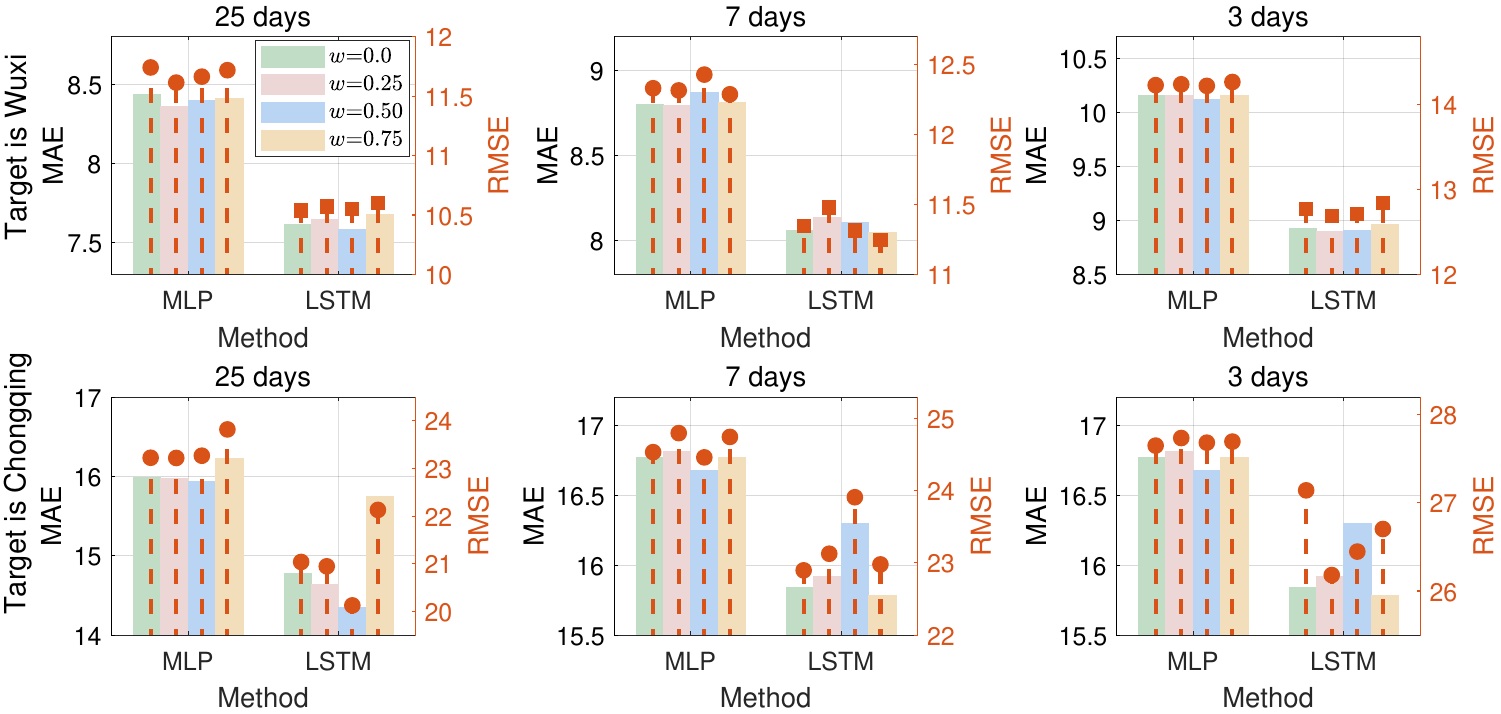}
	\caption{Impact of $w$ on prediction errors.}
	\label{lu13}
\end{figure}

The impact of different balance coefficients on prediction results was analyzed as shown in Figure~\ref{lu13}. It can be observed that the balance coefficient gives the difference between feature embeddings and error loss, and its effectiveness varies with different training data lengths. As the length of training decreases, the optimal balance coefficient tends to increase. A balance coefficient of 0.5 is preferable when the training has a length of 25 days. At this point, both prediction loss and feature embedding loss are equally important. When the training data length is 7 days, a value of 0.75 is preferable. At this point, feature embedding loss becomes more important, and the objective is to minimize the difference in feature embeddings while referencing the features already learned from the source city. Generally, when there is less training data, a larger balance coefficient should be chosen. However, when there is more training data, a balance should be maintained between prediction loss and feature embedding loss.
\subsubsection{Influence of Embedding Dimension}
\begin{figure}[ht]
	\centering
	\includegraphics[width=11cm]{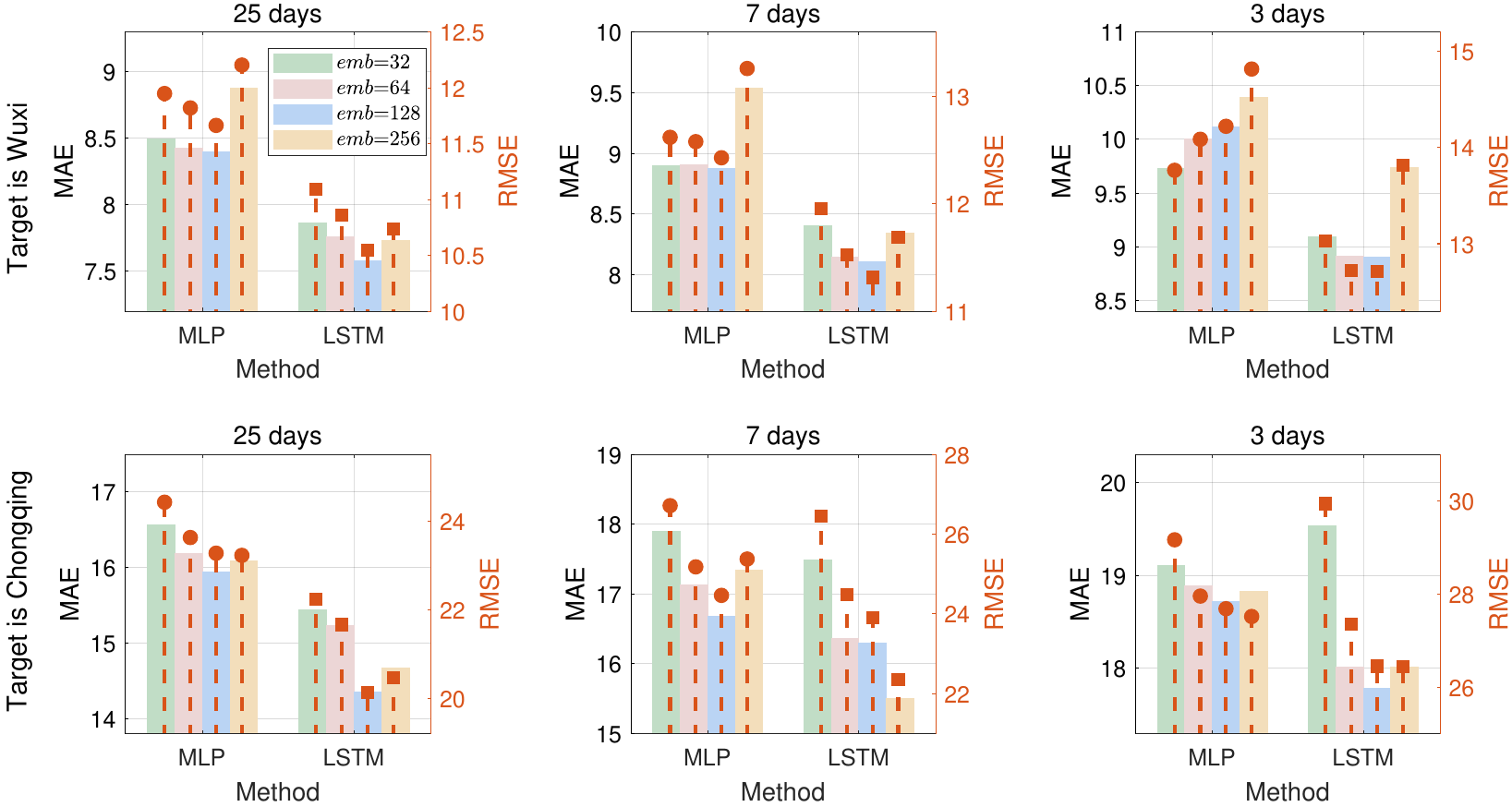}
	\caption{Impact of embedding dimensions on prediction errors.}
	\label{lu14}
\end{figure}

Selecting the appropriate embedding dimension is crucial for extracting input features. As shown in Figure~\ref{lu14}, prediction results for different embedding dimension demonstrate that the prediction error decreases as the embedding dimension increases. However, a larger embedding dimension is not necessarily better, as it may result in challenges in training optimization. Therefore, the embedding dimension should be chosen with consideration of the specific dimensions of the input data. In this study, the dimensions of the input data are 35, encompassing 6 dimensions for passenger flow feature input, 11 dimensions for dynamic covariate input, and 18 dimensions for static covariate input. The optimal feature embedding dimension in our instance is 128.
\subsubsection{Influence of feature network number}
\begin{figure}[ht]
	\centering
	\includegraphics[width=11cm]{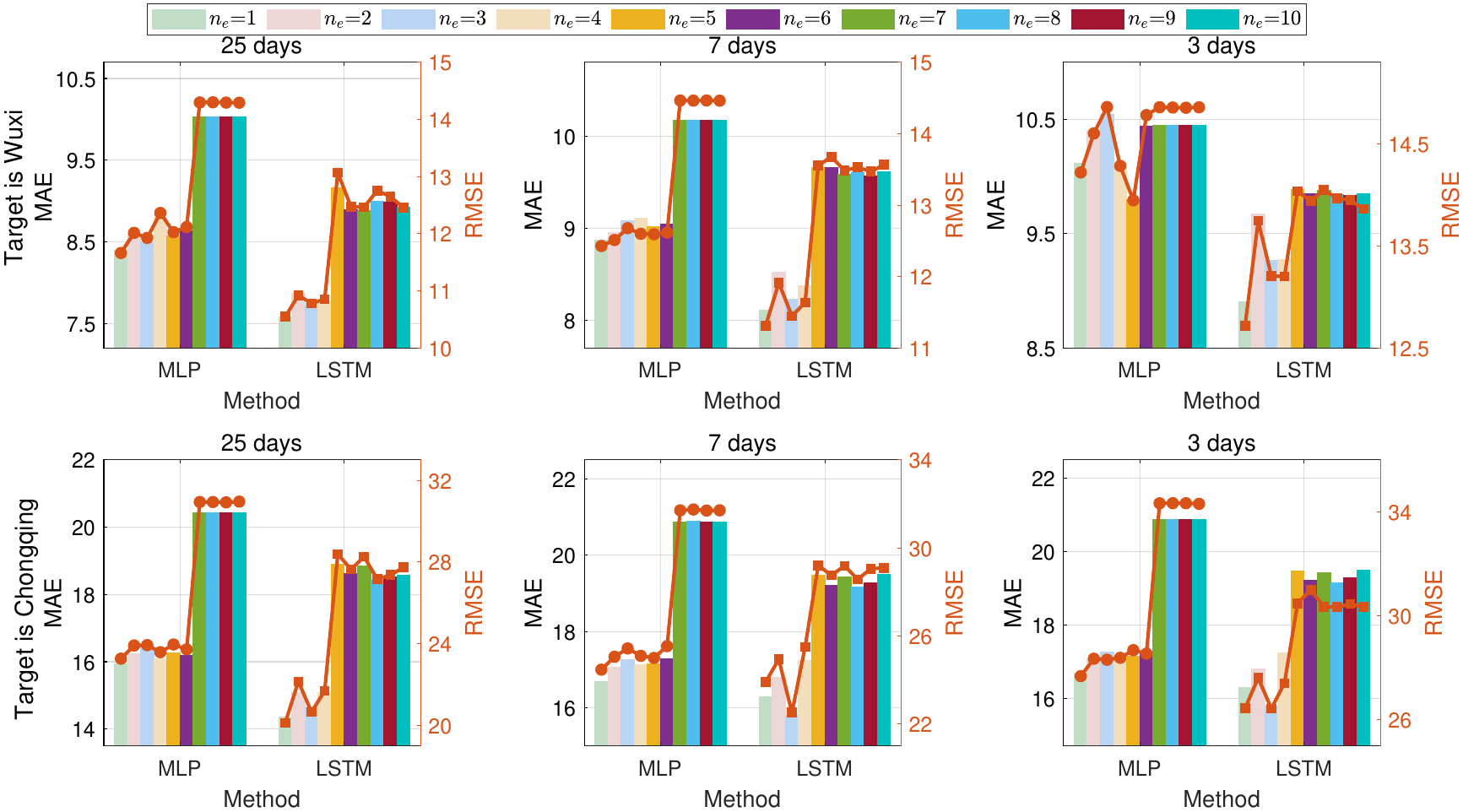}
	\caption{Impact of $n_e$ on prediction errors.}
	\label{lu15}
\end{figure}

The quantity of feature networks within the Encoder-Decoder module is vital in the process of feature extraction. The results of prediction metrics for different network quantities are depicted in Figure~\ref{lu15}. An increase in the quantity of network layers leads to an increase in the prediction error, as observed. Additionally, there is a sudden increase in the prediction metric for the feature networks based on MLP and LSTM at 7 and 5, respectively. Therefore, this implies that a more elaborate feature network has less tolerance for the number of networks. Thus, it is advisable to select a lesser quantity of networks when utilizing an intricate feature network. The determination of the quantity of feature networks in practical application should depend on factors such as input dimensions and the feature network.
\subsubsection{Ablation Studies}
In the final stage, we remove the residual connection (wo Res) and external covariates (wo Ex) in METcross to evaluate their effectiveness. Table~\ref{tab6} illustrates the comparison of the prediction indicators of METcross, wo Ex and wo Res, where the optimal indicator is highlighted in bold black. Results show that METcross provides superior performance compared to wo Ex and wo Res, in which the prediction indicators are slightly decreased. This clarifies that the residual connection and external covariates in METcross has a positive utility. By incorporating external covariates can better extract feature representations between the two cities. In addition, integrating residual connections to merge the base prediction outcomes, it is feasible to reduce prediction errors. 
\begin{table}[ht]
  \scriptsize
  \centering
  \setlength{\tabcolsep}{1.2mm}
  \caption{Prediction metrics for METcross, Wo Ex and Wo Res.}
    \begin{tabular}{cccccccccccccc}
    \toprule
    \multicolumn{2}{c}{Method} & \multicolumn{6}{c}{MLP} & \multicolumn{6}{c}{LSTM} \\
    \midrule
    \multirow{2}{*}{Target} & \multirow{2}{*}{Train data} & \multicolumn{3}{c}{MAE} & \multicolumn{3}{c}{RMSE} & \multicolumn{3}{c}{MAE} & \multicolumn{3}{c}{RMSE} \\
      \cmidrule{3-14}       & & METcross & Wo Ex&Wo Res &METcross & Wo Ex& Wo Res & METcross &Wo Ex&Wo Res&METcross &Wo Ex& Wo Res \\
    \midrule
    {\multirow{3}{*}{WX}} & 25 days & 8.397 & 8.82&\textbf{8.386} & 11.663 & 12.137&\textbf{11.574}   & \textbf{7.58} & 7.803& 7.76  & \textbf{10.548} & 10.847&10.784 \\
     \cmidrule{2-14} & 7 days & \textbf{8.872} & 9.349 &8.973 & \textbf{12.428} &12.731
     & 12.563 &  \textbf{8.106} & 8.394 &8.144 & \textbf{11.315} & 11.759&11.475 \\ 
     \cmidrule{2-14} & 3 days & \textbf{10.117} & 10.636&10.271 & \textbf{14.22} & 14.839
    &14.378 & \textbf{8.903} &9.027& 8.931 & \textbf{12.716} &12.852 &12.783 \\
    \midrule
    {\multirow{3}{*}{CQ}} & 25 days & \textbf{15.935} &16.941& 15.942 & 23.27 &24.692& \textbf{22.938} & \textbf{14.349} &14.726& 14.519 & \textbf{20.124} &20.968& 20.342 \\
     \cmidrule{2-14} & 7 days & \textbf{16.678} &17.094& 16.993 & \textbf{24.465} &25.436& 24.858 & 16.295 &16.755 &\textbf{16.103} & 23.912 &24.513&\textbf{23.483} \\
     \cmidrule{2-14} & 3 days & \textbf{18.718} & 18.85&19.131 & \textbf{27.684} &28.1& 28.213 & \textbf{17.784} & 17.945& 18.298 & \textbf{26.447}& 26.71& 27.286 \\
    \bottomrule
    \end{tabular}
  \label{tab6}
\end{table}
\section{Conclusion}
This study began with constructing the basic and METcross frameworks for cross-city metro passenger flow prediction from the perspective of cross-city data fusion. Subsequently, the constructed frameworks' usability and generalization were evaluated using Chongqing and Wuxi's metro datasets. The results indicate that, compared to the single-city prediction models and the basic framework for multi-city prediction, METcross effectively incorporates data from other cities. Further, an in-depth analysis of METcross and the basic framework results was conducted, providing theoretical references for engineering practice. Finally, ablation experiments were carried out to evaluate the role of external covariates and residual connections in METcross.

The experimental results yielded several key findings: 
\begin{itemize}
\item Metro short-term passenger flow prediction error can be significantly reduced by integrating data from multiple cities. 
\item The overall good predictive performance of the METcross framework is supported by the experimental outcomes from different datasets. 
\item External covariates and residual connection is confirmed to be a valuable component of the METcross framework, according to ablation experiments. 
\item Specified input dimensions and training data lengths should determine the selection of hyperparameters. 
\item By simply fusing the predicted values of two cities, the PF method enhances the utility of the prediction model. Thus, providing insights into cross-city data fusion for established prediction models. Specifically, running prediction models in different cities and merging the predicted values can reduce prediction errors. 
\item The TF approach is an alternative means of improving the utility of established models. Pretraining in the source city and fine-tuning in the target city are sufficient.
\end{itemize}

The experimental results suggest that the multi-city data fusion model concept can also be applied to other traffic prediction tasks, including traffic volume and speed prediction. Furthermore, it can be employed in time series prediction tasks, such as forecasting residential electricity demand. For instance, combining electricity demand data from different cities in the modeling process can aid in predicting the residential electricity demand of a city.

One limitation of this study is that it solely addressed data fusion from two cities. Therefore, it would be advantageous for future research to explore the construction of data from multiple source cities and model them in an integrated fashion. Additionally, incorporating previously developed prediction models for single cities into cross-city fusion is a worthwhile research area. Moreover, future research should consider  integrating graph neural networks to enhance modeling capability, with multiple variables, and broaden its application domain. Hopefully, this study will offer insights for future research in related fields.

\section*{CRediT authorship contribution statement}
\textbf{Wenbo Lu:} Conceptualization, Methodology, Software, Validation, Formal analysis, Writing – original draft, Writing – review \& editing, Supervision. \textbf{Jinhua Xu:} Conceptualization, Resources, Data curation, Writing – review \& editing. \textbf{Peikun Li:} Resources, Data curation, Writing – review \& editing. \textbf{Ting Wang:}  Writing – review \& editing. \textbf{Yong Zhang:} Conceptualization, Methodology, Validation, Writing – original draft, Writing – review \& editing, Supervision.
\section*{Declaration of competing interest}
The authors declare that they have no known competing financial interests or personal relationships that could have appeared to influence the work reported in this paper.
\section*{Acknowledgments}
The authors are grateful to the editor and the reviewers for the valuable comments. This work was supported in part by the National Natural Science Foundation of China under Grant 72071041 and the Jiangsu Province Key R\&D Program under Grant BE2021067.
\section*{Data availability}
Data will be made available on request.
\bibliography{interactcadsample}
\bibliographystyle{cas-model2-names}

\end{document}